\documentclass[prb,amsmath,amssymb,superscriptaddress,twocolumn]{revtex4}
\usepackage{graphicx}
\usepackage{bm}
\usepackage{bbm}
\usepackage{bbold}
\newcommand{\al}{\alpha}
\newcommand{\be}{\beta}

\newcommand{\br}{{\bf r}}

\newcommand{\bk}{{\bf k}}

\newcommand{\bn}{{\bf n}}

\newcommand{\bp}{{\bf p}}

\newcommand{\bK}{{\bf K}}
\newcommand{\bQ}{{\bf Q}}

\newcommand{\eps}{\epsilon}

\newcommand{\si}{\sigma}
\newcommand{\bs}{{\bf \sigma}}
\DeclareMathAlphabet{\mathpzc}{OT1}{pzc}{m}{it} 
\begin{document}
\title{Unconventional superconductivity in a two-dimensional repulsive gas of fermions with spin-orbit coupling}
\author{Luyang Wang}
\author{Oskar Vafek}
\affiliation{National High Magnetic Field Laboratory and Department
of Physics, Florida State University, Tallahassee, Florida 32306,
USA}
\address{}

\date{\today}
\begin{abstract}
We investigate the superconducting instability of a two-dimensional repulsive fermion gas with Rashba spin-orbit coupling $\al_R$. Using renormalization group approach, we find the superconducting transition temperature as a function of the dimensionless ratio $\Theta=\frac{1}{2}m\al_R^2/E_F$ where $E_F=0$ when the smaller Fermi surface shrinks to a (Dirac) point. The general trend is that superconductivity is enhanced as $\Theta$ increases, but in an intermediate regime $\Theta\sim0.1$, a dome-like behavior appears. At a very small value of $\Theta$, the angular momentum channel $j_z$ in which superconductivity occurs is quite high. With increasing $\Theta$, $j_z$ decreases with a step of 2 down to $j_z=6$, after which we find the sequence $j_z=6, 4, 6, 2$, the last value of which continues to $\Theta\rightarrow\infty$. In an extended range of $\Theta$, the superconducting gap predominantly resides on the large Fermi surface, while Josephson coupling induces a much smaller gap on the small Fermi surface. Below the superconducting transition temperature, we apply mean field theory to derive the self-consistent equations and find the condensation energies. The state with the lowest condensation energy is an unconventional superconducting state which breaks time reversal symmetry, and in which singlet and triplet pairings are mixed. In general, these states are topologically nontrivial, and the Chern number of the state with total angular momentum $j_z$ is $C=2j_z$.
\end{abstract}
\maketitle

\section{introduction}
Unconventional superconductivity arising from purely repulsive fermion interactions was first studied by Kohn and Luttinger\cite{KohnLuttinger}. Although the bare interaction is repulsive, screening effects can give rise to attraction between fermions. In three dimensions, $p$-wave superconductivity is found at second order in the interaction\cite{KohnLuttinger}, while for a strictly parabolic dispersion in two dimensions (2D), it is known that repulsive interaction does not induce superconductivity to second order in the interaction, and one has to go to third order for the occurrence of superconductivity\cite{ChubukovPRB}. Since spin-orbit coupling plays an important role in many condensed matter systems, such as topological insulators, noncentrosymmetric systems, and some oxide interfaces, it is natural to ask: what is the role of spin-orbit coupling in this process? Does it enhance superconductivity? And what is the nature of the superconducting state? In this paper, we investigate the unconventional superconductivity in two-dimensional (2D) repulsive Fermi gas with Rashba spin-orbit coupling. (The superconductivity of Rashba model with attractive interaction has been addressed in Ref.\cite{Edelstein,GorkovRashba,MineevSigrist}.)

In an earlier Brief Report\cite{VafekWang}, we have reported some main results of this work. In this paper, we include the details of the calculations. In addition, we clarify the pairing symmetry and topological properties of the unconventional superconducting states we found.

In Rashba model, the strength of the spin-orbit coupling is characterized by the parameter $\al_R$, which is tunable by the application of an external electric field perpendicular to the 2D system. We treat the Rashba spin-orbit coupling $\al_R$ non-perturbatively, so we can analyze the relative values of the mean-field transition temperature $T_c$ for an arbitrary value of the dimensionless ratio $\Theta=\frac{1}{2}m\al_R^2/E_F$, where $m$ is the (bare) fermion mass and $E_F$ is the Fermi energy, measured from the Dirac point. In the strictest sense, in 2D Kosterlitz-Thouless theory should be used to treat the phase transition, and the transition temperature $T_{KT}<T_c$. However, since we are working in the weak coupling limit, the pairing energy scale is much smaller than the zero temperature phase stiffness energy and $1-T_{KT}/T_c\sim T_c/E_F\ll1$, justifying the approach presented here.

Our study is formulated within the renormalization group (RG) approach\cite{Shankar}. We integrate out high energy modes, and derive effective interactions for low energy modes. We perturbatively calculate the renormalization of the interactions, and derive the RG flow equations which describe how the interactions evolve with lowering the energy. The effective interactions, as well as the RG equations, can be decoupled in angular momentum channels. Although singlet and triplet pairs are mixed by spin-orbit coupling\cite{GorkovRashba}, since the Hamiltonian commutes with the $z$-component of the total angular momentum, $J_z=L_z+S_z$, we can label the pair states according to $j_z$, the eigenvalue of $J_z$. The decoupled effective interactions in each channel follow the same RG equation, but have different initial values. In some channels, they diverge at some energy scale as the RG flow runs. Among all the channels, the highest energy scale at which the divergence occurs is identified with the superconducting transition temperature.

The Fermi surface splits into two due to spin-orbit coupling, a large one and a small one, with helicity $\lambda=+1$ and -1, respectively. We find that the superconducting gap residing on the large Fermi surface always dominates, while momentum space Josephson coupling induces superconductivity on the small Fermi surface. The superconductivity is enhanced by spin-orbit coupling, since now it appears at second order of the repulsive interaction instead of third order. With $\Theta$ increasing from small values to infinity, the angular momentum channel $j_z$ in which Cooper pairs condense decreases as a arithmetic sequence with step 2, with an exception in the intermediate range of $\Theta$ (see Fig. \ref{fig:phase diag}). In the limit of large $\Theta$, we find $j_z=2$. Our results can also be derived diagrammatically by summing the leading logarithms to all orders in perturbation theory, as has been done traditionally in treating Kohn-Luttinger effect\cite{GorkovMelikBarkhudarov,ChubukovReview}. Also, our approach is similar to that of Ref.\cite{RaghuKivelsonScalapino} (see also\cite{RaghuKivelson2011}), which implements a two-step RG by first eliminating high energy modes down to an artificial cutoff and then running the RG flow from the cutoff. However, our single step RG is more economical.

In the superconducting state, mean field theory is applied to find the self-consistent equations and the condensation energies. There are two solutions to the self-consistent equations, one fully gaps the Fermi surfaces and breaks time reversal symmetry (TRS), and the other has gap nodes and does not break TRS. The former has a lower condensation energy, hence is the physical state. In this state, only one of the two $\pm j_z$ pairing components is finite, and singlet and triplet pairings are mixed. For example, $j_z=2$ state is a mixture of $d_{x^2-y^2}+id_{xy}$ singlets, $p_x+ip_y$ spin-up triplets and $f_{x^3-3xy^2}+if_{3x^2y-y^3}$ spin-down triplets. These TRS breaking states are topologically nontrivial, with the Chern number $C=2j_z$.

\begin{figure}[h]
\includegraphics[width=0.4\textwidth]{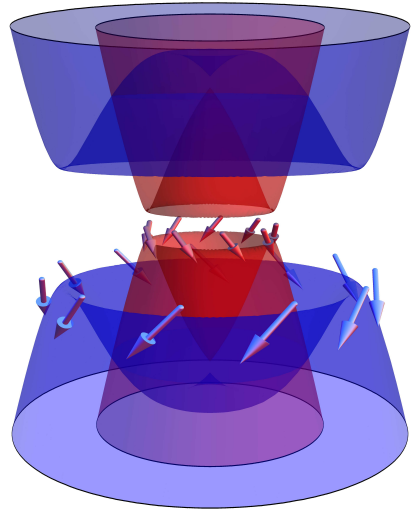}
\caption{A schematic plot of $\vec{\mathcal{D}}_\lambda$ (without the phase factor) around the two gapped Fermi surfaces. }\label{fig:d-vector}
\end{figure}

It is convenient to define a three-component vector $\vec{\mathcal{D}}_\lambda$ in such a way that the gap function on helicity-$\lambda$ Fermi surface is $(\vec{\mathcal{D}}_\lambda\cdot\vec{\Sigma})(i\sigma_y)$, where $\vec{\Sigma}=(\sigma_x,\sigma_y,\mathbb{1})$. For a general noncentrosymmetric superconductor, the gap function is usually defined as $(\psi\mathbb{1}+\vec{d}\cdot\vec{\sigma})(i\sigma_y)$， where $\psi$ is the order parameter for spin-singlet pairing while $\vec{d}$ is that for spin-triplet pairing. In our case, the $z$-component of the $\vec{d}$-vector is zero since all the triplets are polarized. So we combine the $x$- and $y$-component of $\vec{d}$ with $\psi$ to form the new vector $\vec{\mathcal{D}}_\lambda$. In this way the gap function can be represented by $\vec{\mathcal{D}}_\lambda$, which can be viewed graphically. We find
\begin{eqnarray}\label{eq:D-vector}
\vec{\mathcal{D}}_\lambda=\Delta_\lambda i\lambda e^{ij_z\theta_\bk}(\sin\theta_\bk,-\cos\theta_\bk,-\lambda),
\end{eqnarray}
where $\Delta_\lambda$ is the pairing amplitude on the helicity-$\lambda$ Fermi surface.
We plot $\vec{\mathcal{D}}_\pm$ (without the phase factor) around the two gapped Fermi surfaces schematically in Fig.\ref{fig:d-vector}.

This paper is organized as follows. In Sec.\ref{Hamil}, we set up the Hamiltonian and solve the eigenenergies and eigenstates for the noninteracting Hamiltonian. In Sec.\ref{pertur}, we formulate the problem using path integrals, and perturbatively expand the action to second order. Two diagrams, the particle-particle bubble and particle-hole bubble, contribute to the renormalization of the interaction. In Sec.\ref{Pi}, we explicitly calculate the particle-hole bubble, which will show up in our final expression for the superconducting transition temperature. In Sec.\ref{high}, the higher order expansion is calculated, to serve as the RG flow. In Sec.\ref{RG}, RG approach is applied, and the decoupled flow equations are found and solved in each angular momentum channel. In Sec.\ref{evalu}, the effective couplings and superconducting transition temperature $T_c$ are computed. The symmetry and topological properties of the unconventional superconducting states are illustrated in Sec.\ref{Superconduting states}. We summarize the paper in Sec.\ref{summary}. The mean field theory below $T_c$, including the Ginzburg-Landau theory, is derived in Appendix \ref{mft}.
\section{Hamiltonian}\label{Hamil}
We start from the single particle Hamiltonian of a two-dimensional Fermi gas with spin-orbit coupling,
\begin{eqnarray}
H=H_0+H_{SO}+H_{int}
\end{eqnarray}
where the free electron term is
\begin{eqnarray}
H_0&=&\frac{\bk^2}{2m},
\end{eqnarray}
the spin-orbit coupling term is
\begin{eqnarray}
H_{SO}&=&\al_R(\si\times\bk)\cdot\hat{\bn}\nonumber\\
&=&\al_R(\bs_x k_y-\si_y k_x)\nonumber\\
&=&\al_R k\left(\begin{array}{cc}0&ie^{-i\theta_\bk}\\-ie^{i\theta_\bk}&0\end{array}\right),
\end{eqnarray}
where $\si$'s are Pauli matrices, and the interacting term $H_{int}$ is dealt with later. Here $\theta_\bk$ is the angle between $\bk$ and $k_x$-axis. The eigenenergies of the non-interacting Hamiltonian are
\begin{eqnarray}
\eps_{\bk\lambda}&=&\frac{k^2}{2m}-\lambda\al_R k=\frac{(k-\lambda k_R)^2}{2m}-\frac{k_R^2}{2m},
\end{eqnarray}
and the corresponding eigenstates are
\begin{eqnarray}
\eta_{\bk\lambda}=\frac{1}{\sqrt{2}}\left(\begin{array}{cc}1\\i\lambda e^{i\theta_\bk}\end{array}\right)
\end{eqnarray}
where $k_R=m\al_R$, $\lambda=\pm1$ and $k_\lambda=k_F+\lambda k_R$ with $k_F=\sqrt{2mE_F+k_R^2}=\sqrt{k_F^{(0)2}+k_R^2}$. The Fermi surface is split into two two Fermi surfaces by spin-orbit coupling, with different helicities. In second quantization form,
\begin{eqnarray}
H_0&=&\sum_{\bk\lambda}\frac{k^2}{2m} c^\dagger_{\bk\lambda}c_{\bk\lambda},\\
H_{SO}&=&\al_R\sum_{\bk\al\be}c^\dagger_{\bk\al}[(\bs_{\al\be}\times\bk)\cdot\hat{\bn}]c_{\bk\be},
\end{eqnarray}
where $c^\dagger_{\bk\lambda}$ and $c_{\bk\lambda}$ creates and annihilates fermions with spin $\lambda=\uparrow$ or $\downarrow$ and momentum $\bk$, respectively. Diagonalized in the helicity basis, the kinetic Hamiltonian becomes
\begin{eqnarray}
H_{kin}&=&H_0+H_{SO}\nonumber\\
&=&\sum_{\bk}\left(c_{\bk\uparrow}^\dagger, c_{\bk\downarrow}^\dagger\right)\left(\begin{array}{cc}\frac{k^2}{2m}&\al_R kie^{i\theta_\bk}\\-\al_R kie^{i\theta_\bk} &\frac{k^2}{2m}\end{array}\right)\left(\begin{array}cc_{\bk\uparrow}\\c_{\bk\downarrow}\end{array}\right)\nonumber\\
&=&\sum_{\bk\lambda}\eps_{\bk\lambda}a^\dagger_{\bk\lambda} a_{\bk\lambda}
\end{eqnarray}
where $a^\dagger_{\bk\lambda}$ and $a_{\bk\lambda}$ creates and annihilates fermions with helicity $\lambda=1$ or $-1$, and momentum $\bk$, respectively. We will only focus on the general case with $\eps_{\bk\lambda}>0$, and will not discuss the case with $\eps_{\bk\lambda}<0$. for $\lambda=\pm 1$,The density of states of the two bands are
\begin{eqnarray}
N_{\pm}(\eps)=\nu_{2D}\left(1\pm\frac{k_R}{\sqrt{k_R^2+2m\eps}}\right)
\end{eqnarray}
Here $\nu_{2D}=\frac{m}{2\pi}$ is the density of states per spin in 2D for $\al_R=0$. The total density of states at any energy is $2\nu_{2D}$, the same as that of a 2D free electron gas, as expected.
For simplification, we assume that the interaction between fermions is short range repulsive interaction, instead of Coulomb interaction. The interacting Hamiltonian reads
\begin{eqnarray}
H_{int}&=&\frac{u}{2}\frac{1}{L^2}\sum_{\bk_1...\bk_4}\sum_{\si\si'}\delta_{\bk_1+\bk_2,\bk_3+\bk_4}
c_{\bk_1\si}^\dagger c_{\bk_2\si'}^\dagger c_{\bk_3\si'}c_{\bk_4\si},\nonumber\\
\end{eqnarray}
where $u>0$. The components of $\bk$ belong to the Born-von Karman set $\{2\pi n/L\}$ where $n$ is an integer and $L$ is the linear size of the system. In the weak coupling limit, $u\nu_{2D}\ll 1$.
In terms of the helicity eigenmodes, the interacting Hamiltonian is written as
\begin{eqnarray}
H_{int}&=&\frac{u}{2L^2}\sum_{\bk_1...\bk_4}\sum_{\mu\nu\lambda\rho}\delta_{\bk_1+\bk_2,\bk_3+\bk_4}\nonumber\\ &\times&[\eta_{\bk_1,\mu},\eta_{\bk_4,\rho}] [\eta_{\bk_2,\nu},\eta_{\bk_3,\lambda}] a_{\bk_1\mu}^\dagger a_{\bk_2\nu}^\dagger a_{\bk_3\lambda}a_{\bk_4\rho},
\end{eqnarray}
where the scalar product of two spinors is
\begin{eqnarray}
[\eta_{\bk\lambda},\eta_{\bk'\lambda'}]=\frac{1}{2}[1+\lambda\lambda' e^{-i(\theta_\bk-\theta_{\bk'})}].
\end{eqnarray}
Antisymmetrizing the interaction, we have
\begin{widetext}
\begin{eqnarray}
H_{int}&=&\frac{u}{16L^2}\sum_{\bk_1...\bk_4}\sum_{\mu\nu\lambda\rho}\delta_{\bk_1+\bk_2,\bk_3+\bk4}(\mu e^{-i\theta_{\bk_1}}-\nu e^{-i\theta_{\bk_2}})(\rho e^{i\theta_{\bk_4}}-\lambda e^{i\theta_{\bk_3}}) a_{\bk_1\mu}^\dagger a_{\bk_2\nu}^\dagger a_{\bk_3\lambda}a_{\bk_4\rho}.
\end{eqnarray}
\end{widetext}

\section{Perturbative expansion to second order}\label{pertur}
\begin{figure}[h]
\begin{center}
\begin{tabular}{cc}
\includegraphics[width=0.25\textwidth]{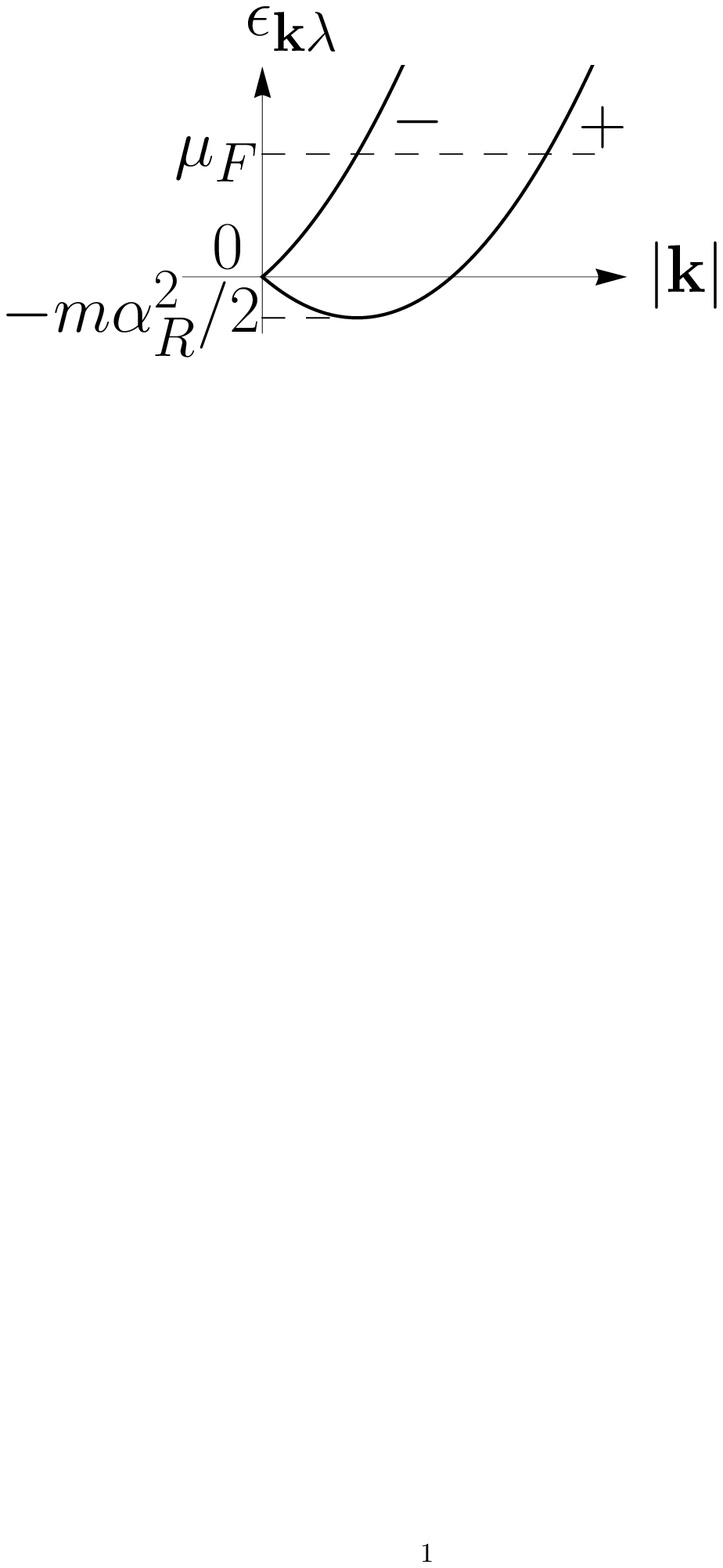}&
\hspace{.06\textwidth}
\includegraphics[width=0.15\textwidth]{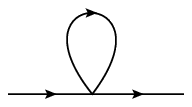}
\end{tabular}
\end{center}
 \caption{(Left) The dispersion relation. (Right) First order (tadpole) correction to self-energy.
  }\label{fig:disp}
\end{figure}

In the perturbative expansion, we integrate out the high energy modes between the energy cutoff $A$ and $\Omega\ll A$ about the two Fermi surfaces, and derive an effective interaction for low energy modes. Suppose that the noninteracting action is $S_0$ and the interacting action is $S_{int}$. Let the action be expressed as follows:
\begin{eqnarray}
S(\phi_<,\phi_>)&=&S_0(\phi_<)+S_0(\phi_>)+S_{int}(\phi_<,\phi_>)
\end{eqnarray}
where $\phi_<$ and $\phi_>$ represent low energy modes and high energy modes, respectively, and $S_0$ is a quadratic function of its arguments that separates into low energy and high energy pieces and $S_{int}$ is quartic, which mixes the two. The partition function is
\begin{eqnarray}
Z=\int D\phi e^{-S_0-S_{int}}
=\int D\phi_< e^{-S_0(\phi_<)-S_{int}'(\phi_<)}.
\end{eqnarray}
Cumulant expansion gives
\begin{eqnarray}
-S_{int}'&=&-\langle S_{int}\rangle+\frac{1}{2}(\langle S_{int}^2\rangle-\langle S_{int}\rangle^2)+...
\end{eqnarray}
to the second order, where $\langle$ $\rangle$ denotes averages with respect to the high energy modes with action $S_0$.
Now we return to the full Hamiltonian
\begin{widetext}
\begin{eqnarray}
H=\sum_{\bk\lambda}\eps_{\bk\lambda}a^\dagger_{\bk\lambda} a_{\bk\lambda}+\frac{u}{16L^2}\sum_{\bk_1...\bk_4}\sum_{\mu\nu\lambda\rho}\delta_{\bk_1+\bk_2,\bk_3+\bk4}(\mu e^{-i\theta_{\bk_1}}-\nu e^{-i\theta_{\bk_2}})(\rho e^{i\theta_{\bk_4}}-\lambda e^{i\theta_{\bk_3}}) a_{\bk_1\mu}^\dagger a_{\bk_2\nu}^\dagger a_{\bk_3\lambda}a_{\bk_4\rho},
\end{eqnarray}
\end{widetext}
then the partition function is
\begin{eqnarray}
Z=\int Da^*_+a_+a^*_-a_- e^{-S_0-S_{int}},
\end{eqnarray}
where
\begin{eqnarray}
S_0&=&\int_0^\be
d\tau\sum_{\bk,\lambda=\pm}a^*_{\bk\lambda}(\tau)(\frac{\partial}{\partial\tau}+\eps_{\bk\lambda}-\mu_F) a_{\bk\lambda}(\tau),\\
S_{int}&=&\int_0^\be d\tau\sum_{1,2,3,4}U(1,2,3,4)a^*(1)a^*(2)a(3)a(4).
\end{eqnarray}
In the above expressions, $\be=1/(k_BT)$, and $\mu_F$ is the exact chemical potential which acts to preserve average particle density. We adopt a shorthand expression for the multiple summations $\sum_{1,2,3,4}(...)=\int_0^\be d\tau_1...d\tau_4\sum_{\bk_1\bk_2\bk_3\bk_4}\sum_{\mu\nu\lambda\rho}(...)$,
\begin{eqnarray}
U(1,2,3,4)&=&\frac{u}{16L^2}\int_0^\be d\tau\prod_{j=1}^4\delta(\tau-\tau_j)\delta_{\bk_1+\bk_2,\bk_3+\bk_4}\nonumber\\
&\times&(\mu e^{-i\theta_{\bk_1}}-\nu e^{-i\theta_{\bk_2}})(\rho e^{i\theta_{\bk_4}}-\lambda e^{i\theta_{\bk_3}})
\end{eqnarray}
and $a(j)=a_{\bk_j\al_j}(\tau_j)$, where $\al_j=\{\mu,\nu,\lambda,\rho\}$.
At first order, the effective interaction is just the bare interaction. In the Cooper channel we have
\begin{eqnarray}\label{eq:bare}
-\langle S_{int}\rangle&=&-\frac{u}{4L^2}\sum_{\bk,\bk'}\int_0^\be d\tau\sum_{\mu\lambda}\mu\lambda e^{i(\theta_{\bk'}-\theta_\bk)}\nonumber\\
&\times&a^*_{\bk\mu}(\tau)a^*_{-\bk\mu}(\tau)a_{-\bk'\lambda}(\tau)a_{\bk'\lambda}(\tau)
\end{eqnarray}
Another term at first order is the tadpole diagram, as shown in Fig.\ref{fig:disp}, which gives a correction to the chemical potential. The correction is $\delta \mu_F=-\frac{1}{2}u(\langle\hat{\rho}_+\rangle+\langle\hat{\rho}_-\rangle)$ where $\hat{\rho}_\pm=\int\frac{d^2\bk}{(2\pi)^2}a^\dagger_{\bk\pm}a_{\bk\pm}$. Such {\it negative} interaction correction must be absorbed in the chemical potential counterterm, $\mu_F-\mu^{(0)}_F=\frac{1}{2}u(\langle\hat{\rho}_{+}\rangle+\langle\hat{\rho}_{-}\rangle)+\mathcal{O}(u^2)$, which is {\it positive}, and which guarantees that the average particle density remains fixed. In general, we are not aware of any argument why interactions should not renormalize the areas of the individual Fermi surfaces, while of course maintaining their sum fixed, but to first order we find no such renormalization.
At second order, the term $\langle S_{int}\rangle^2$ cancel out unconnected diagrams in $\langle S_{int}^2\rangle$, and we are left with connected diagrams, including particle-hole and particle-particle diagrams, which renormalize the effective interaction in the Cooper channel. Following Shankar's notation\cite{Shankar}, we have
\begin{widetext}
\begin{eqnarray}
-\delta S_{int}&=&\frac{1}{2}\langle S_{int}^2\rangle_{\mbox{con}}=\frac{1}{2}\int_0^\be d\tau d\tau'\sum_{4321}\sum_{4'3'2'1'}
U(4321)U(4'3'2'1')\langle a^*(4)a^*(3)a(2)a(1)a^*(4')a^*(3')a(2')a(1')\rangle\nonumber\\
&=&\mbox{ZS}+\mbox{ZS'}+\mbox{BCS},
\end{eqnarray}
\end{widetext}
where both ZS and ZS' contribute to the particle-hole bubble.
\begin{figure}[h]
\begin{center}
\includegraphics[width=0.25\textwidth]{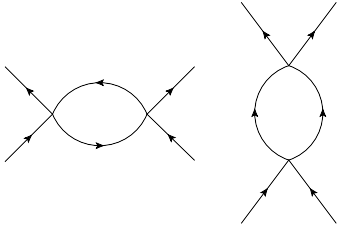}
\end{center}
 \caption{Second order correction to 4-pt scattering amplitude.}\label{fig:2nd}
\end{figure}
The particle-hole and particle-particle bubbles are shown in Fig.\ref{fig:2nd}, with expressions
\begin{widetext}
\begin{eqnarray}
\mbox{ZS}&=&4\int_0^\be d\tau d\tau'\sum_{4321}\sum_{4'3'2'1'} U(4'41'2)U(3'32'1)G(4',2')G(1',3')a^*(4)a^*(3)a(2)a(1),\\
\mbox{ZS'}&=&-4\int_0^\be d\tau d\tau'\sum_{4321}\sum_{4'3'2'1'} U(4'41'1)U(3'32'2)G(4',2')G(1',3')a^*(4)a^*(3)a(2)a(1),\\
\mbox{BCS}&=&2\int_0^\be d\tau d\tau'\sum_{4321}\sum_{4'3'2'1'} U(432'1')U(4'3'21)G(1',4')G(2',3')a^*(4)a^*(3)a(2)a(1),
\end{eqnarray}
\end{widetext}
where the Green's functions are, for example
\begin{eqnarray}
G(1,2)=\delta_{i_1,i_2}\delta_{\bk_1,\bk_2}G_{\bk_1i_1}(\tau_1-\tau_2)
\end{eqnarray}
where $i_j$ indicates the helicity of particle $j$. Note that ZS' term differs from ZS term by exchanging two incoming (or equivalently, two outgoing) particles, with a minus sign which results from Fermi statistics.
\subsection{Particle-Hole Bubble}
Now we need to evaluate the diagrams. The Kronecker delta and the momentum conservation facilitate the simplification of the expressions,
\begin{widetext}
\begin{eqnarray}
\mbox{ZS}&=&\frac{u^2}{64L^4}\int_0^\be d\tau d\tau'\sum_{i_4i_3i_2i_1}\sum_{\al\be}\sum_{\bk_4\bk_3\bk_2\bk_1\bK} \delta_{\bk_4+\bk_3,\bk_2+\bk_1}G_{\bK\al}(\tau-\tau')G_{\bK+\bk_1-\bk_3\be}(\tau-\tau')\nonumber\\
&\times&(\al e^{-i\theta_{\bK}}-i_{4}e^{-i\theta_{\bk_4}})(i_{2}e^{i\theta_{\bk_2}}-\be e^{i\theta_{\bK+\bk_1-\bk_3}})
(\be e^{-i\theta_{\bK+\bk_1-\bk_3}}-i_{3}e^{-i\theta_{\bk_3}})(i_{1}e^{i\theta_{\bk_1}}-\al e^{i\theta_{\bK}})\nonumber\\ &\times&a^*_{\bk_4i_4}(\tau)a^*_{\bk_3i_3}(\tau')a_{\bk_2i_2}(\tau)a_{\bk_1i_1}(\tau').
\end{eqnarray}
\end{widetext}
In the Cooper channel we have $\bk_4=-\bk_3=\bk$, $\bk_1=-\bk_2=\bk'$, $i_4=i_3=\mu$ and $i_2=i_1=\lambda$. We can also set the imaginary times on all Grassman terms to $\tau$. Therefore,
\begin{eqnarray}
\mbox{ZS}&=&\frac{u^2}{64L^2}\int_0^\be d\tau\sum_{\mu\lambda}\sum_{\bk,\bk'}\Pi_{\mu\lambda}^{(+)}(\bk,\bk')\nonumber\\
&\times& a^*_{\bk\mu}(\tau)a^*_{-\bk\mu}(\tau)a_{-\bk'\lambda}(\tau)a_{\bk'\lambda}(\tau),
\end{eqnarray}
where
\begin{eqnarray}
\Pi_{\mu\lambda}^{(+)}(\bk,\bk')&=&\frac{1}{\be}\sum_{\omega_n} \int\frac{d^2\bp}{(2\pi)^2}\sum_{\al\be}G_{\bp\al}(i\omega_n)G_{\bp+\bk'+\bk\be}(i\omega_n)\nonumber\\
&\times&(\al e^{-i\theta_{\bp}}-\mu e^{-i\theta_{\bk}})(\lambda e^{i\theta_{-\bk'}}-\be e^{i\theta_{\bp+\bk'+\bk}})\nonumber\\
&\times&(\be e^{-i\theta_{\bp+\bk'+\bk}}-\mu e^{-i\theta_{-\bk}})(\lambda e^{i\theta_{\bk'}}-\al e^{i\theta_{\bp}})
\end{eqnarray}
As mentioned, the only difference between ZS and ZS' is the exchange of two incoming or outgoing particles in the interaction and a minus sign in front. Thus we have
\begin{eqnarray}
\mbox{ZS'}&=&-\frac{u^2}{64L^2}\int_0^\be d\tau\sum_{\mu\lambda}\sum_{\bk,\bk'}\Pi_{\mu\lambda}^{(-)}(\bk,\bk')\nonumber\\ &\times& a^*_{\bk\mu}(\tau)a^*_{-\bk\mu}(\tau)a_{-\bk'\lambda}(\tau)a_{\bk'\lambda}(\tau)
\end{eqnarray}
where
\begin{eqnarray}
\Pi_{\mu\lambda}^{(-)}(\bk,\bk')&=&\Pi_{\mu\lambda}^{(+)}(\bk,-\bk')=\Pi_{\mu\lambda}^{(+)}(-\bk,\bk').
\end{eqnarray}
Combining ZS and ZS' terms, we find
\begin{eqnarray}
\mbox{ZS+ZS'}&=&-\frac{u^2}{64L^2}\int_0^\be d\tau\sum_{\mu\lambda}\sum_{\bk,\bk'}(\Pi_{\mu\lambda}(\bk,\bk')-\Pi_{\mu\lambda}(-\bk,\bk'))\nonumber\\
&\times& a^*_{\bk\mu}(\tau)a^*_{-\bk\mu}(\tau)a_{-\bk'\lambda}(\tau)a_{\bk'\lambda}(\tau)
\end{eqnarray}
where we have set
\begin{eqnarray}
\Pi_{\mu\lambda}(\bk,\bk')=\Pi_{\mu\lambda}^{(-)}(\bk,\bk').
\end{eqnarray}
Thus the term multiplying the four Grassman numbers in the Cooper channel is automatically odd under both $\bk\rightarrow-\bk$ and $\bk'\rightarrow-\bk'$. After the frequency sum, $\Pi_{\mu\lambda}(\bk,\bk')$ becomes
\begin{eqnarray}
\Pi_{\mu\lambda}(\bk,\bk')
&=&\sum_{\al\be}\int\frac{d^2\bp}{(2\pi)^2}\frac{n_F(\eps_{\bp\al})-n_F(\eps_{\bp+\bk-\bk'\be})} {\eps_{\bp\al}-\eps_{\bp+\bk-\bk'\be}} \nonumber\\
&\times& F_{\mu\lambda\al\be}(\bk,\bk',\bp)
\end{eqnarray}
where the phase factor $F_{\mu\lambda\al\be}(\bk,\bk',\bp)$ is
\begin{eqnarray}
F_{\mu\lambda\al\be}(\bk,\bk',\bp)&=&(\al e^{-i\theta_{\bp}}-\mu e^{-i\theta_{\bk}})
(\lambda e^{i\theta_{\bk'}}-\be e^{i\theta_{\bp+\bk-\bk'}})\nonumber\\
&\times&(\be e^{-i\theta_{\bp+\bk-\bk'}}-\mu e^{-i\theta_{-\bk}})(\lambda e^{i\theta_{-\bk'}}-\al e^{i\theta_{\bp}}).\nonumber\\
\end{eqnarray}
Later, we will show that $F_{\mu\lambda\al\be}(\bk,\bk',\bp)$ is a real function times the complex factor $e^{i(\theta_{\bk'}-\theta_{\bk})}$, which is expected since the bare interaction is of such form (see Eq.(\ref{eq:bare})), and the renormalization of $u$ should be real.
\subsection{Particle-Particle Bubble}
Now we turn to the particle-particle bubble - the BCS term.
In the Cooper channel, we have
\begin{eqnarray}
\mbox{BCS}&=&\frac{u^2}{8L^2}\int_0^\be d\tau\sum_{\mu\lambda}\sum_{\bk\bk'}\mu\lambda e^{i(\theta_{\bk'}-\theta_\bk)}P(\Omega)\nonumber\\
&\times& a^*_{\bk\mu}(\tau)a^*_{-\bk\mu}(\tau)a_{-\bk'\lambda}(\tau)a_{\bk'\lambda}(\tau),
\end{eqnarray}
where
\begin{eqnarray}
P(\Omega)&=&\frac{1}{\be}\sum_{\omega_n}\int\frac{d^2\bp}{(2\pi)^2}\sum_{\al} G_{\bp\al}(i\omega_n)G_{-\bp\al}(-i\omega_n)\nonumber\\
&=&(N_++N_-)\ln\frac{A}{\Omega}.
\end{eqnarray}
In the above expression, $N_{\pm}$ are the densities of states of the two bands at the Fermi energy. The BCS term is the lowest-order term which gives rise to a logarithm. In the case with attractive electron-electron interaction, this term will result in the superconducting instability. However, in our case, where the interaction between electrons is repulsive, to search for the superconducting instability we need to go to higher order terms with logarithms.

To second order, the full correction to the action can be written as
\begin{eqnarray}
\delta S_{int}&=&\frac{u^2}{64L^2}\int_{0}^\be d\tau\sum_{\bk\bk'}\sum_{\mu\lambda}V_{\mu\lambda}(\bk,\bk')\nonumber\\ &\times&a^*_{\bk\mu}(\tau)a^*_{-\bk\mu}(\tau)a_{-\bk'\lambda}(\tau)a_{\bk'\lambda}(\tau),
\end{eqnarray}
where we write
\begin{eqnarray}
V_{\mu\lambda}(\bk,\bk')=V^{ph}_{\mu\lambda}(\bk,\bk')+V^{pp}_{\mu\lambda}(\bk,\bk'),
\end{eqnarray}
in which
\begin{eqnarray}
V^{ph}_{\mu\lambda}(\bk,\bk')&=&\Pi_{\mu\lambda}(\bk,\bk')-\Pi_{\mu\lambda}(-\bk,\bk'),\\
V^{pp}_{\mu\lambda}(\bk,\bk')&=&-8\mu\lambda e^{i(\theta_{\bk'}-\theta_\bk)}(N_+ +N_-)\ln\frac{A}{\Omega}.
\end{eqnarray}
\section{Evaluation of $\Pi_{\mu\lambda}(\bk,\bk')$}\label{Pi}
\subsection{Phase Factor in $\Pi_{\mu\lambda}(\bk,\bk')$}
We now calculate the phase factor $F_{\mu\lambda\al\be}(\bk,\bk',\bp)$ appearing in $\Pi_{\mu\lambda}(\bk,\bk')$. We expect that $\Pi_{\mu\lambda}(\bk,\bk')$ is a function of $\theta_\bk-\theta_{\bk'}$ while not a function of $\theta_\bk$ and $\theta_{\bk'}$ separately, because of the rotational invariance of the Fermi surfaces. Also, as mentioned, $\Pi_{\mu\lambda}(\bk,\bk')$ should be a real function multiplying the complex factor $e^{i(\theta_{\bk'}-\theta_{\bk})}$. Before seeing this clearly in the expression of $\Pi_{\mu\lambda}(\bk,\bk')$, we need some algebra. Let $\bk-\bk'=\bQ$, then
\begin{eqnarray}
F_{\mu\lambda\al\be}(\bk,\bk',\bp)&=&(\al e^{-i\theta_{\bp}}-\mu e^{-i\theta_{\bk}})(\lambda e^{i\theta_{\bk'}}-\be e^{i\theta_{\bp+\bQ}})\nonumber\\
&\times&(\be e^{-i\theta_{\bp+\bQ}}-\mu e^{-i\theta_{-\bk}})(\lambda e^{i\theta_{-\bk'}}-\al e^{i\theta_{\bp}}).\nonumber\\
\end{eqnarray}
Let $\phi=\theta_{\bk'}-\theta_\bk$. Using
\begin{eqnarray}
e^{i\theta_{\bp+\bQ}}&=&\frac{pe^{i\theta_\bp}+Qe^{i\theta_\bQ}}{|\bp+\bQ|},\\
e^{i\theta_\bQ}&=&e^{i\theta_{\bk-\bk'}}=\frac{ke^{i\theta_\bk}-k'e^{i\theta_{\bk'}}}{|\bk-\bk'|},
\end{eqnarray}
and shifting $\theta_\bp$ to $\theta_\bp+\theta_\bQ$, we have
\begin{widetext}
\begin{eqnarray}
\frac{F_{\mu\lambda\al\be}(\bk,\bk',\bp)}{e^{i\phi}}&=& 2(\cos\phi-\mu\lambda)+2\mu\lambda\al\be\frac{p+Q\cos{\theta_{\bp}}}{|\bp+\bQ|} -2\al\be\left(\frac{((k^2+k'^2)\cos\phi-2kk')(p\cos2\theta_\bp+Q\cos\theta_\bp)} {(\bk-\bk')^2|\bp+\bQ|}\right)\nonumber\\
&-&(\mu+\lambda)\left(\frac{(k+k')(1-\cos\phi)}{|\bk-\bk'|} \left(\be\frac{p\cos\theta_\bp+Q}{|\bp+\bQ|}-\al \cos\theta_\bp\right)\right)\nonumber\\ &+&(\mu-\lambda)\left(\frac{(k-k')(\cos\phi+1)}{|\bk-\bk'|}\left(\be\frac{p\cos\theta_\bp+Q}{|\bp+\bQ|}- \al\cos\theta_\bp\right) \right),
\end{eqnarray}
\end{widetext}
where
\begin{eqnarray}
Q&=&|\bk-\bk'|=\sqrt{k^2+k'^2-2kk'\cos\phi},\\
|\bp+\bQ|&=&\sqrt{p^2+Q^2+2pQ\cos\theta_\bp}.
\end{eqnarray}
We have neglected the terms containing $\sin\theta_\bp$, since
\begin{eqnarray}
\int_0^{2\pi}d\theta f(\cos\theta)\sin\theta=-\int_{-\pi}^{\pi}d\theta f(-\cos\theta)\sin\theta=0.
\end{eqnarray}

It is natural to rescale all the momenta by the "Rashba momentum" $k_R=m\al_R$, then
\begin{eqnarray}\label{eq:rescale}
\eps_{\bk\mu}&=&\frac{k_R^2}{2m}(\bk^2-2\mu k).
\end{eqnarray}
Similarly, the Fermi energy can be rescaled as
\begin{eqnarray}
\eps_F&=&\frac{E_F}{k_R^2/2m}=k_{F\mu}^2-2\mu k_{F\mu},
\end{eqnarray}
so the rescaled Fermi momenta are, in terms of the rescaled Fermi energy,
\begin{eqnarray}
k_{F\mu}=\sqrt{\eps_F+1}+\mu.
\end{eqnarray}
The zero temperature occupation factors require either $p<k_{F\al}$ and $|\bp+\bQ|>k_{F\be}$, or $p>k_{F\al}$ and $|\bp+\bQ|<k_{F\be}$.
Using the above rescaling, we can write $\Pi_{\mu\lambda}(\bk,\bk')$ as
\begin{eqnarray}
\Pi_{\mu\lambda}(\bk,\bk')=e^{i\phi}2m\Lambda_{\mu\lambda}(\Theta,\cos\phi)
\end{eqnarray}
where $\Theta$ is defined as $\Theta=\frac{1}{2}m\al_R^2/E_F=1/\eps_F$, and $\Lambda_{\mu\lambda}(\Theta,\cos\phi)$ is real and equals
\begin{eqnarray}\label{eq:Lambda_polar}
&&\Lambda_{\mu\lambda}(\Theta,\cos\phi)=\sum_{\al\be}\int_0^{\infty}\frac{dpp}{(2\pi)^2}\int_{-\pi}^\pi d\theta_\bp\nonumber\\ &\times&\frac{\Theta(k_{F\al}-p)-\Theta(k_{F\be}-|\bp+\bQ|)}{(p-\al)^2-(|\bp+\bQ|-\be)^2} \frac{F_{\mu\lambda\al\be}(\bk,\bk',\bp)}{e^{i\phi}}.
\end{eqnarray}
$\bk$ and $\bk'$ reside on Fermi surface $\mu$ and $\lambda$ respectively, so $k=k_{F\mu}$ and $k'=k_{F\lambda}$.
Such form of $\Pi_{\mu\lambda}(\bk,\bk')$ suggests that the particle-hole contribution in the Cooper channel at second order can be rewritten as
\begin{eqnarray}
V^{ph}_{\mu\lambda}(\bk,\bk')=e^{i\phi}2m[\Lambda_{\mu\lambda}(\Theta,\cos\phi) +\Lambda_{\mu\lambda}(\Theta,\cos(\phi+\pi))]
\end{eqnarray}
and since $\cos(\phi+\pi)=-\cos\phi$, the Taylor expansion of the term in the square brackets contains only even powers of $\cos\phi$. As a result, it can be decoupled into even angular momentum channels
\begin{eqnarray}
V^{ph}_{\mu\lambda}(\bk,\bk')=e^{i\phi}4m\sum_{j_z=0,2,4...}V_{\mu\lambda}^{(j_z)}(\Theta)\cos{j_z\phi},
\end{eqnarray}
where the Fourier transform reads
\begin{eqnarray}
V_{\mu\lambda}^{(j_z)}(\Theta)&=&\frac{1}{2\pi}\int_0^{2\pi}d\phi e^{-ij_z\phi}\nonumber\\
&\times&[\Lambda_{\mu\lambda}(\Theta,\cos\phi)+\Lambda_{\mu\lambda}(\Theta,-\cos\phi)]
\end{eqnarray}
where $V_{\mu\lambda}^{(j_z)}(\Theta)$ is real for all $j_z$.
\subsection{Evaluation of $\Lambda_{\mu\lambda}$ in Elliptic Coordinates}
To evaluate $\Lambda_{\mu\lambda}$, we need to calculate the double integral in Eq.(\ref{eq:Lambda_polar}). We notice that neither the angular nor the radial part can be done analytically. But numerical computation of the double integral is time consuming. Therefore, to proceed, we choose to rewrite $\Lambda_{\mu\lambda}$ in elliptic coordinates. In order to do so we need to shift $\bp$ to $\bp-\frac{1}{2}\bQ$ and transform from the polar coordinates to elliptic coordinates, $x\in[1,\infty),\psi\in[0,2\pi)$ by substituting
\begin{eqnarray}
p_\parallel&=&\frac{1}{2}Qx\cos\psi,\\
p_\perp&=&\frac{1}{2}Q\sqrt{x^2-1}\sin\psi.
\end{eqnarray}
Using
\begin{eqnarray}
|\bp\pm\frac{1}{2}\bQ|=\frac{1}{2}Q(x\pm\cos\psi)
\end{eqnarray}
and the Jacobian
\begin{eqnarray}
J(\frac{p_\parallel,p_\perp}{\psi,x})=\frac{Q^2}{4}\frac{x^2-\cos^2\psi}{\sqrt{x^2-1}},
\end{eqnarray}
and changing the variables to $y=\cos\psi$ since $\psi$ appears only as $\cos\psi$, we have
\begin{widetext}
\begin{eqnarray}
\Lambda_{\mu\lambda}(\Theta,\cos\phi)&=&
\frac{1}{2}\frac{Q}{(2\pi)^2}\sum_{\al\be}\int_{1}^{-1}\frac{dy}{\sqrt{1-y^2}}\int_1^\infty\frac{dx}{\sqrt{x^2-1}} \left[\frac{-\delta_{\al,\be}}{(Qx-2\al)y}+\frac{-\delta_{\al,-\be}}{(Qx+2\al)x}\right]\nonumber\\
&\times&\left[\Theta(\frac{2k_{F\al}}{Q}+y-x)-\Theta(\frac{2k_{F\be}}{Q}-y-x)\right]\nonumber\\
&\times&\left[2(\cos\phi-\mu\lambda)(x^2-y^2)-4\al\be\left(\frac{(k_{F\mu}^2+k_{F\lambda}^2)\cos\phi-2k_{F\mu} k_{F\lambda}}{Q^2} \right) \left(x^2y^2-\frac{1}{2}(x^2+y^2)\right)\right.\nonumber\\
&+&2\al\be\mu\lambda\left(x^2+y^2-2\right)\nonumber\\
&+&(\mu+\lambda)\frac{k_{F\mu}+k_{F\lambda}}{Q}(1-\cos\phi)\left((\al-\be)(x^2-1)y-(\al+\be)x(1-y^2)\right) \nonumber\\
&-&\left.(\mu-\lambda)\frac{k_{F\mu}-k_{F\lambda}}{Q}(1+\cos\phi)\left((\al-\be)(x^2-1)y-(\al+\be)x(1-y^2)\right) \right].
\end{eqnarray}
For $\al=\be$, we perform the $y$-integral first, which can be done in terms of elementary functions; similarly, for $\al=-\be$, we perform the $x$-integral first. The remaining integral needs to be done numerically. The step functions impose the upper and lower limit on the integrals. The final result for the antisymmetrized combination $\Lambda_{\mu\lambda}^{(S)}(\Theta,\cos\phi)=\frac{1}{2}(\Lambda_{\mu\lambda}(\Theta,\cos\phi)+ \Lambda_{\mu\lambda}(\Theta,-\cos\phi))$ is shown in Fig.\ref{fig:Lambda}.

\begin{figure}[h]
\begin{center}
\begin{tabular}{cc}
\includegraphics[width=0.34\textwidth]{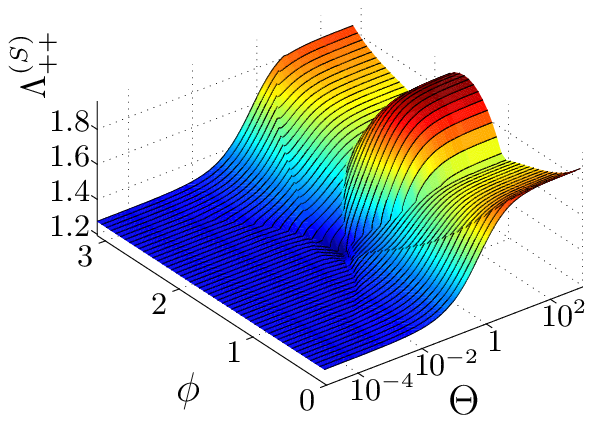}
\includegraphics[width=0.34\textwidth]{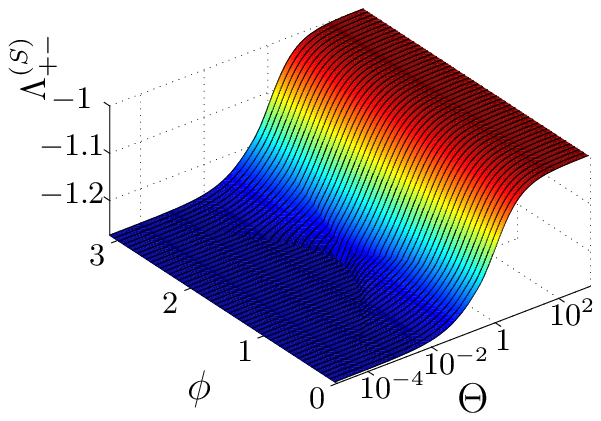}
\end{tabular}
\end{center}
 \caption{Relative angle $\phi=\theta_{\bk'}-\theta_{\bk}$ and $\Theta=\frac{1}{2}m\alpha^2_R/E_F$ dependence of the interaction function $\Lambda^{(S)}_{\mu\lambda}$.
 $\Lambda^{(S)}_{++}$ (left) and $\Lambda^{(S)}_{+-}$ (right) start from
 $\pm\frac{4}{\pi}$ at $\Theta=0$ and develop $\phi$ dependence for
 finite $\Theta$, while $\Lambda^{(S)}_{--}$ remains $\frac{4}{\pi}$
 for any $\Theta$.
 }\label{fig:Lambda}
\end{figure}
\end{widetext}

\section{Perturbative Expansion to Higher Order}\label{high}
\begin{figure}[ht]
\begin{center}
\includegraphics[width=0.3\textwidth]{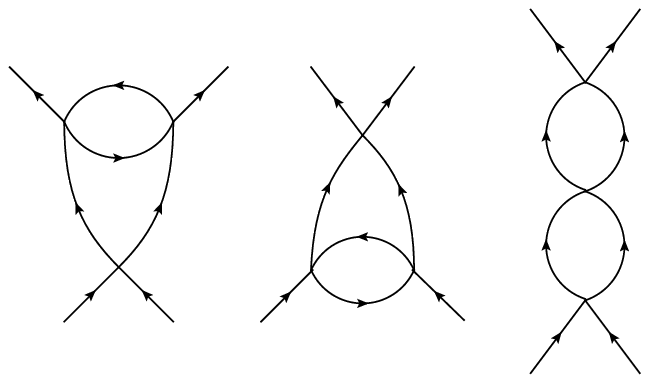}
\hspace{0.1\textwidth}
\includegraphics[width=0.45\textwidth]{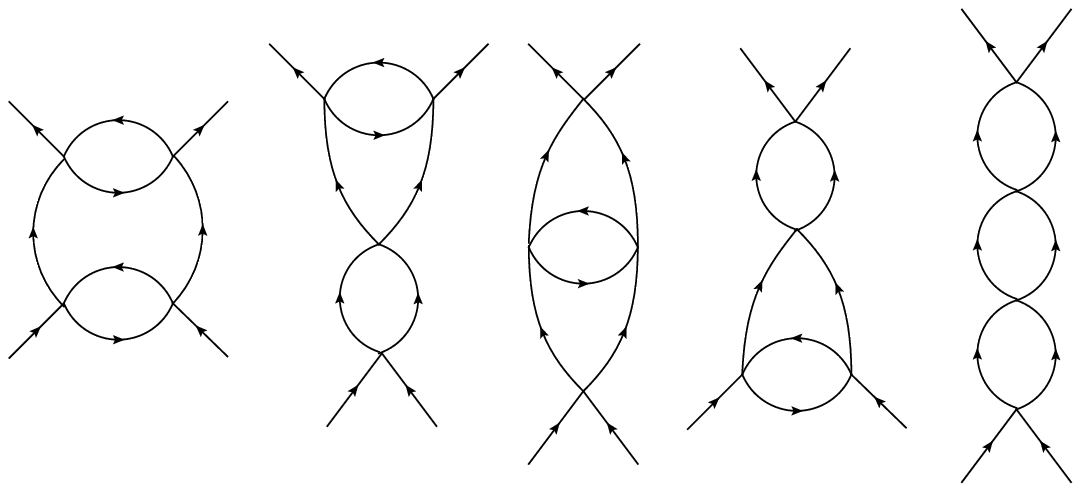}
\end{center}
  \caption{(Left) 3$^{rd}$ order correction to the 4-pt scattering
  amplitude.
  (Right) 4$^{th}$ order correction. We display only the diagrams which contain
  logarithmic enhancement.
  }\label{fig:graphs}
\end{figure}

Now we consider the 3rd and 4th order terms which renormalize the Cooper channel. These terms are represented by diagrams shown in Fig.\ref{fig:graphs}, and used to derive the renormalization group (RG) equations governing the flows of Cooper channel couplings. At third order of the cumulant expansion, we have three terms with logarithms,
\begin{widetext}
\begin{eqnarray}
&&\int_0^\be d\tau\sum_{\mu\lambda}\sum_{\bk\bk'} e^{i\phi}\left\{\frac{u^3}{2^7L^2}\ln\frac{A}{\Omega}\left[\sum_{\al}\mu\al N_{\al}2mV^{(0)}_{\al\lambda}+\sum_{\al}\lambda\al N_{\al}2mV^{(0)}_{\mu\al}\right]-\frac{u^3}{16L^2}\mu\lambda(N_++N_-)^2\ln^2\frac{A}{\Omega}\right\}\nonumber\\ &\times&a^*_{\bk\mu}(\tau)a^*_{-\bk\mu}(\tau)a_{-\bk'\lambda}(\tau)a_{\bk'\lambda}(\tau),
\end{eqnarray}
\end{widetext}
while at fourth order, we have five such terms,
\begin{widetext}
\begin{eqnarray}
&&\int_0^\be d\tau\sum_{\mu\lambda}\sum_{\bk\bk'} e^{i\phi}\left\{-\frac{u^4}{2^8L^2}\ln^2\frac{A}{\Omega}\left[(N_++N_-)\sum_{\al}\mu\al N_{\al}2mV^{(0)}_{\al\lambda}+(N_++N_-)\sum_{\al}\lambda\al N_{\al}2mV^{(0)}_{\mu\al}\right.\right.\nonumber\\
&+&\left.\lambda\mu\sum_{\al\be} \al N_{\al}\be N_\be 2mV^{(0)}_{\al\be}\right] +\frac{u^4}{2^{11} L^2}\ln\frac{A}{\Omega}\sum_{j_z=0,\pm2,\pm4...}e^{ij_z\phi}\left[2mV^{(j_z)}_{\mu\al}2mV^{(j_z)}_{\al\lambda}\right] \nonumber\\
&+&\frac{u^4}{32L^2}\left.\mu\lambda(N_++N_-)^3\ln^3\frac{A}{\Omega}\right\} a^*_{\bk\mu}(\tau)a^*_{-\bk\mu}(\tau)a_{-\bk'\lambda}(\tau)a_{\bk'\lambda}(\tau).
\end{eqnarray}
\end{widetext}
\section{Renormalization Group Approach}\label{RG}
We project the renormalized coupling into angular momentum channels denoted by $j_z$, and define $V_{\mu\lambda}^{r(j_z)}$ through the expression
\begin{eqnarray}
S_{int}'&=&\frac{1}{L^2}\int_0^\be d\tau\sum_{\bk\bk'}\sum_{\mu\lambda}e^{i\phi}\sum_{j_z}e^{ij_z\phi}\nonumber\\ &\times&V_{\mu\lambda}^{r(j_z)}a^*_{\bk\mu}(\tau)a^*_{-\bk\mu}(\tau)a_{-\bk'\lambda}(\tau)a_{\bk'\lambda}(\tau).
\end{eqnarray}
From the perturbative expansion to 4th order, we see that $V_{\mu\lambda}^{r(j_z\neq0)}$ contains only the terms of even powers in $u$, while $V_{\mu\lambda}^{r(0)}$ contains terms of all powers in $u$.
\subsection{$j_z\neq0$}
At $j_z\neq0$, only the second and fourth order terms contribute to the effective coupling in Cooper channel,
\begin{eqnarray}
V_{\mu\lambda}^{r(j_z)}&=&\frac{u^2m}{2^5}V_{\mu\lambda}^{(j_z)}\nonumber\\
&-&\frac{u^4m^2}{2^9}\sum_\al N_\al V_{\mu\al}^{(j_z)}V_{\al\lambda}^{(j_z)}\ln\frac{A}{\Omega}+...
\end{eqnarray}
where ... represents terms of order $u^4$ which do not contain (large) logarithm as well as of higher order in $u$. If we define a dimensionless coupling matrix $g_{\mu\lambda}^{(j_z)}=\frac{1}{2^5}u^2m\sqrt{N_\mu N_\lambda}V_{\mu\lambda}^{(j_z)}$, then the above equation can be written in a matrix form,
\begin{eqnarray}
g^{r(j_z)}&=&g^{(j_z)}-2g^{(j_z)}*g^{(j_z)}\ln\frac{A}{\Omega},
\end{eqnarray}
where "$*$" represents the matrix multiplication. Taking the logarithmic derivative of the right hand side, then to, and including, $\mathcal{O}(u^4)$, we have the RG flow equation
\begin{eqnarray}
\frac{dg^{r(j_z)}}{d\ln\Omega}=2g^{r(j_z)}*g^{r(j_z)}
\end{eqnarray}
Then $g^{r(j_z)}$ is diagonalized by a unitary transformation,
\begin{eqnarray}
\frac{d}{d\ln\Omega}(Ug^{r(j_z)} U^\dagger)=2Ug^{r(j_z)} U^\dagger Ug^{r(j_z)} U^\dagger
\end{eqnarray}
after which the eigenvalues of $g^{r(j_z)}$, $g^{r(j_z)}_\pm$, satisfy the RG equation separately,
\begin{eqnarray}\label{eq:RG}
\frac{d}{d\ln\Omega}(2g^{r(j_z)}_{\pm})=(2g^{r(j_z)}_{\pm})^2.
\end{eqnarray}
The matrix $g^{(j_z)}$ can be written as, in terms of its elements,
\begin{eqnarray}
g^{(j_z)}=\frac{1}{2}(g^{(j_z)}_{++}+g^{(j_z)}_{--})\mathbb{1} +\frac{1}{2}(g^{(j_z)}_{++}-g^{(j_z)}_{--})\sigma_z+g^{(j_z)}_{+-} \sigma_x,
\end{eqnarray}
where $\mathbb{1}$ is the identity matrix. The eigenvalues of $g^{(j_z)}$ are
\begin{eqnarray}
g^{(j_z)}_{\pm}=\frac{1}{2}(g^{(j_z)}_{++}+g^{(j_z)}_{--})\pm\sqrt{\frac{1}{4}(g^{(j_z)}_{++}-g^{(j_z)}_{--})^2 +g^{(j_z)2}_{+-}}.
\end{eqnarray}
Then the RG equation (\ref{eq:RG}) is readily integrated, which yields
\begin{eqnarray}\label{eq:gpm}
g_{\pm}^{r(j_z)}(\Omega)=\frac{g_{\pm}^{(j_z)}}{1+2g_{\pm}^{(j_z)}\ln\frac{A}{\Omega}},
\end{eqnarray}
where the initial eigenvalues of ${g^{r}}^{(j_z)}_{\mu\lambda}|_{\Omega=A}$, for $j_z\neq0$, are
\begin{eqnarray}\label{eq:gell pm}
g^{(j_z)}_{\pm}&=&\frac{u^2m}{2^5}\left(\frac{1}{2}(N_+V^{(j_z)}_{++}+N_-V^{(j_z)}_{--})\right. \nonumber\\ &\pm&\left.\sqrt{\frac{1}{4}(N_+V^{(j_z)}_{++}-N_-V^{(j_z)}_{--})^2+N_+N_-{V_{+-}^{(j_z)}}^2}\right).
\end{eqnarray}
The density of states on the two Fermi surfaces are $N_{\pm}=\nu_{2D}\left(1\pm\frac{\sqrt{\Theta}}{\sqrt{1+\Theta}}\right)$.
If $g^{(j_z)}_{\pm}<0$ for some $j_z$ and $\Theta$, then the associated renormalized coupling (\ref{eq:gpm}) diverges at a scale
\begin{eqnarray}\label{eq:Tc}
T_c^{(j_z)}\sim\Omega^{*(j_z)}=Ae^{-\frac{1}{|g^{(j_z)}_{eff,\pm}|}}
\end{eqnarray}
where $g^{(j_z)}_{eff,\pm}=2g^{(j_z)}_{\pm}$. While the assignment between $T_c$ and $\Omega^*$ cannot reliably determine the prefactor of the exponential term, the relative dependence on $\al_R$ is in the exponential factor, which we can determine. This allows us to compare the dependence of the ratio of superconducting transition temperatures on $\al_R$.
\subsection{$j_z=0$}
For $j_z=0$, the renormalized coupling is
\begin{widetext}
\begin{eqnarray}
V_{\mu\lambda}^{r(0)}&=&\frac{u}{4}\mu\lambda+(\frac{u}{8})^2 2mV^{(0)}_{\mu\lambda}-\frac{u^2}{8}\mu\lambda(N_++N_-)\ln\frac{A}{\Omega} +\frac{u^3}{16}\mu\lambda(N_++N_-)^2\ln^2\frac{A}{\Omega}\nonumber\\
&-&\frac{2u^3}{16^2}\ln\frac{A}{\Omega}2m(\mu\sum_\al \al N_\al V^{(0)}_{\al\lambda}+\lambda\sum_\al \al N_\al V^{(0)}_{\al\mu})-\frac{u^4}{32}\mu\lambda(N_++N_-)^3\ln^3\frac{A}{\Omega}\nonumber\\
&+&\frac{u^4}{16^2}\ln^2\frac{A}{\Omega}2m\left[(N_++N_-)(\mu\sum_\al N_\al V^{(0)}_{\al\lambda}+\lambda\sum_\al \al N_\al V^{(0)}_{\al\mu})+\lambda\mu\sum_{\al\be}\al\be N_\al N_\be V^{(0)}_{\al\be}\right]\nonumber\\
&-&\frac{2u^4}{16^3}\ln\frac{A}{\Omega}\sum_\al N_\al (2mV^{(0)}_{\mu\al}2mV^{(0)}_{\al\lambda}).
\end{eqnarray}
\end{widetext}
To the 4th order, this can be written in a matrix form
\begin{eqnarray}
g^{r(0)}&=&g^{(0)}-2g^{(0)}*g^{(0)}\ln\frac{A}{\Omega}+4g^{(0)}*g^{(0)}*g^{(0)}\ln^2\frac{A}{\Omega}\nonumber\\ &-&8g^{(0)}*g^{(0)}*g^{(0)}*g^{(0)}\ln^3\frac{A}{\Omega},
\end{eqnarray}
where the elements of $g^{(0)}$ are
\begin{eqnarray}\label{eq:initial}
g_{\mu\lambda}^{(0)}=\frac{u}{4}\mu\lambda\sqrt{N_\mu N_\lambda}+\frac{u^2m}{2^5}\sqrt{N_\mu N_\lambda}V^{(0)}_{\mu\lambda}.
\end{eqnarray}
Diagonalizing $g^{r(0)}$ and assuming the eigenvalues are $g^{r(0)}_\pm$, we have
\begin{eqnarray}
g^{r(0)}_\pm&=&g^{(0)}_\pm-2g^{(0)2}_\pm\ln\frac{A}{\Omega}+4g^{(0)3}_\pm\ln^2\frac{A}{\Omega} -8g_\pm^{(0)4}\ln^3\frac{A}{\Omega}\nonumber\\
&\approx&\frac{g^{(0)}_\pm}{1+2g^{(0)}_\pm\ln\frac{A}{\Omega}}.
\end{eqnarray}
Then the RG equation is
\begin{eqnarray}
\frac{dg_\pm^{r(0)}}{d\ln\Omega}&=&\frac{2g^{(0)2}_\pm}{(1+2g^{(0)}_\pm\ln\frac{A}{\Omega})^2}=2g^{r(0)2}_\pm,
\end{eqnarray}
which is the same as Eq.(\ref{eq:RG}), but with a different initial condition (\ref{eq:initial}).
\section{Evaluation of the Effective Couplings and $T_c$}\label{evalu}
\begin{figure}[ht]
\begin{center}
\includegraphics[width=0.475\textwidth]{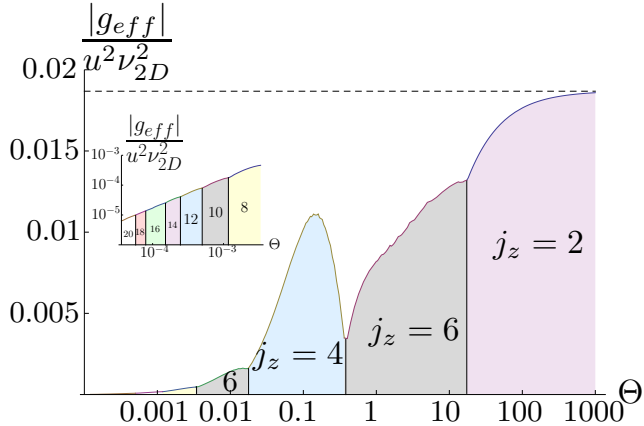}
\end{center}
  \caption{The effective coupling appearing in the expression for
  $T_c\approx Ae^{-1/|g_{eff}|}$ as a function of $\Theta=\frac{1}{2}m\alpha^2_R/E_F$. $\nu_{2D}=\frac{m}{2\pi}$. The dashed
  line at $0.0187$ is the $\Theta\rightarrow\infty$ asymptote.
  }\label{fig:phase diag}
\end{figure}

For each value of $\Theta$, we compare $g_{\pm}^{(j_z)}$ in different angular momentum channels, and the most negative one determines the energy scale at which the superconducting instability occurs and corresponds to the highest $T_c$, which is the physical transition temperature. To within our numerical accuracy, we find that $\Lambda_{--}^{(S)}$ remains $4/\pi$ for any $\Theta$ and has no $\phi$ dependence, thus $V_{--}^{(j_z=0)}=4/\pi$, and $V_{--}^{(j_z\neq0)}=0$ for any $\Theta$. In addition, as can be seen in Fig.\ref{fig:Lambda}, for $\Theta\gtrsim\mathcal{O}(0.01)$ most angle dependence is in $V_{++}^{ph}$, while there is only very weak angle dependence in $V_{+-}^{ph}$. To the first order $\mathcal{O}(u)$, $g_+^{(j_z=0)}>0$ and $g_-^{(j_z=0)}=0$, meaning that no pairing instability occurs. To the second order $\mathcal{O}(u^2)$, we find that $g_{-}^{(j_z=0)}>0$ for any $\Theta$ due to increase in both $V_{++}^{(j_z=0)}$ and $V_{+-}^{(j_z=0)}$, latter of which becomes less negative. Therefore, no superconductivity occurs in $j_z=0$ channel. Since $V_{++}^{ph}$ has most angle dependence, superconductivity resides predominantly on the large Fermi surface and is determined by some $V_{++}^{(j_z)}$ turning negative. In Fig.\ref{fig:phase diag} we show the $\Theta$ dependence of the couplings for the $g_-^{(j_z)}$-channel which has the highest $T_c$. The general trend is that $T_c$ increases with $\Theta$, and the channel in which pairing instability occurs follows a decreasing arithmetic sequence with step 2. At small value of $\Theta$, $T_c$ is small and $j_z$ is very high; while as $\Theta$ increases, $T_c$ increases and $j_z$ decreases. An exception happens at an intermediate range of $\Theta$, starting with $\Theta\sim 0.005$, where we find the sequence $j_z=6,4,6,2$, the last value of which continues to $\Theta\to\infty$, and a dome-like behavior in $T_c$ appears at $\Theta\sim0.1$ in channel $j_z=4$.

\section{Unconventional superconducting states}\label{Superconduting states}
\subsection{Time-reversal symmetry breaking}
Now we need to determine which linear combination of the two possible $\pm j_z$ states has the lowest (most negative) condensation energy below $T_c$. Adopting the arguments of Anderson and Morel that when the most attractive coupling is in $j_z$ channel, the error in the ground state energy involved in neglecting other channels is very small\cite{AndersonMorel1961}, we study this problem below $T_c$ within mean-field theory. The details are presented in Appendix \ref{mft}. We replace the full angular dependence of the original pairing potential with just its projection on the most dominant $j_z$ channel, an approximation which we expect to hold away from the boundaries separating ground states with different angular momentum. The self-consistent mean-field equations are derived, and then solved both at $T=0$ and near $T_c$. We find either a solution which breaks TRS and fully gaps the Fermi surfaces, i.e. only one of the two $\pm j_z$ pairing components is finite, or a solution with equal admixture of $\pm j_z$ and with gap nodes. Comparing their condensation energies we find that the TRS breaking solution is lower by a factor of 1.5 just below $T_c$ and by $e/2\approx 1.36$ at $T=0$. For values of $\Theta\gtrsim 0.005$, the gap on the large Fermi surface is much larger than the gap on the small one due to the smallness of ratio $V_{+-}^{(j_z)}/V_{++}^{(j_z)}$. For smaller value of $\Theta$ the two gaps may be comparable.
\subsection{Pairing symmetry}
Since the pairing occurs between fermions with the same helicity, singlets and triplets are mixed. Under time reversal operation, the creation and annihilation operators transform as $\hat{K}a_{\bk\lambda}=-i\lambda e^{i\theta_\bk}a_{-\bk\lambda}$ and $\hat{K}a^\dagger_{\bk\lambda}=i\lambda e^{-i\theta_\bk}a^\dagger_{-\bk\lambda}$. Therefore, the operator $i\lambda e^{-i\theta_\bk}a^\dagger_{\bk\lambda}a^\dagger_{-\bk\lambda}$ creates a Cooper pair, of which the angular wave function is
\begin{eqnarray}\label{eq:pairing}
&&\frac{1}{2}i\lambda e^{i(j_z-1)\theta_\bk}(|\uparrow\rangle+i\lambda e^{i\theta_\bk}|\downarrow\rangle) (|\uparrow\rangle-i\lambda e^{i\theta_\bk}|\downarrow\rangle)\nonumber\\
&=&\frac{1}{2}i\lambda\left[e^{i(j_z-1)\theta_\bk}|\uparrow\uparrow\rangle+e^{i(j_z+1)\theta_\bk} |\downarrow\downarrow\rangle\right.\nonumber\\
&             &\left.-i\lambda e^{i j_z\theta_\bk}(|\uparrow\downarrow\rangle-|\downarrow\uparrow\rangle)\right].
\end{eqnarray}
When Fourier transformed to real space, $\theta_\bk$ is replaced by $\theta_\br$, the polar angle in the center of mass coordinate system of the Cooper pair. Therefore, the Cooper pair is a coherent superposition of a quarter of spin-up triplet with orbital angular momentum $\ell=j_z-1$, a quarter of spin-down triplet with $\ell=j_z+1$ and a half of singlet with $\ell=j_z$. Because $j_z$ is an even number, the wave function is antisymmetric under the exchange of the two fermions. As seen in Fig.\ref{fig:phase diag}, for large $\Theta$, we have $j_z=2$, which means that the Cooper pair is a mixture of $p_x+ip_y$ spin-up triplet, $d_{x^2-y^2}+id_{xy}$ singlet and $f_{x^3-3xy^2}+if_{3x^2y-y^3}$ spin-down triplet.

As mentioned in the introduction, a three-component vector $\vec{\mathcal{D}}_\lambda$ is defined in such a way that the gap function on helicity-$\lambda$ Fermi surface is $(\vec{\mathcal{D}}_\lambda\cdot\vec{\Sigma})(i\sigma_y)$, where $\vec{\Sigma}=(\sigma_x,\sigma_y,\mathbb{1})$. Comparing this with Eq.(\ref{eq:pairing}), we find
\begin{eqnarray}\label{eq:D-vector}
\vec{\mathcal{D}}_\lambda=\Delta_\lambda i\lambda e^{ij_z\theta_\bk}(\sin\theta_\bk,-\cos\theta_\bk,-\lambda),
\end{eqnarray}
where $\Delta_\lambda$ is the pairing amplitude on the helicity-$\lambda$ Fermi surface.
We plot $\vec{\mathcal{D}}_\pm$ (without the phase factor) around the two Fermi surfaces which, as shown in Appendix \ref{mft}, are fully gapped, schematically in Fig.\ref{fig:d-vector}. The dispersion is given in Eq.(\ref{eq:disp}).

The pairing symmetry can also be seen from the mean field Hamiltonian, which is derived in Eq.(\ref{eq:MFH}). It can be written as $H=\frac{1}{2}\sum_{\bk}\Phi_\bk^\dagger h_0(\bk)\Phi_\bk$, where, if we let $a_{\bk+}=a_\bk$, $ a_{\bk-}=b_{\bk}$, then $\Phi_\bk=(a_\bk,b_{\bk},a_{-\bk}^\dagger,b_{-\bk}^\dagger)^T$, and
\begin{eqnarray}
h_0(\bk)=\left(\begin{array}{cccc}\xi_{\bk+}&0&2\Delta_{a}&0\\ 0&\xi_{\bk-}&0&2\Delta_{b}\\ 2\Delta_{a}^*&0&-\xi_{\bk+}&0\\ 0&2\Delta_{b}^*&0&-\xi_{\bk-}\end{array}\right).
\end{eqnarray}
In the above expression, $\Delta_j$ is defined as
\begin{eqnarray}
\Delta_j&=&\sum_{s=\pm}\Delta_{js}e^{i(sj_z-1)\theta_\bk}
\end{eqnarray}
for $j=a,b$ (see Eq.(\ref{eq:gap})), and $\xi_{\bk\lambda}=\epsilon_{\bk\lambda}-E_F$. We have changed the notation: use $a$ and $b$ to denote the large and small Fermi surfaces, while "$+$" and "$-$" to denote the two components of $\pm j_z$.  In Appendix \ref{mft} we show that only one component of $\Delta_{j\pm}$ is finite in the physical state, corresponding to spontaneous TRS breaking. The two states with either $\Delta_{j+}$ or $\Delta_{j-}$ vanishing have the same energy. Assume $\Delta_{j+}$ is finite, then $\Delta_j=\Delta_{j+}e^{i(j_z-1)\theta_\bk}$ for $j=a,b$. Here, $\Delta_{j+}$ with $j=a,b$ corresponds to $\Delta_\lambda$ with $\lambda=\pm1$ in Eq.(\ref{eq:D-vector}), respectively.
After a unitary transformation, the Hamiltonian is transformed back to spin basis,
\begin{eqnarray}
H=\frac{1}{2}\sum_{\bk}\Psi_\bk^\dagger h(\bk)\Psi_\bk,
\end{eqnarray}
where $\Psi_\bk=(c_{\bk\uparrow},c_{\bk\downarrow},c^\dagger_{-\bk\uparrow},c^\dagger_{-\bk\downarrow})^T$, and
\begin{widetext}
\begin{eqnarray}
h(\bk)=\left(\begin{array}{cccc}\xi_{\bk}&\al_R(k_y+ik_x)&\Delta_{t}e^{i(j_z-1)\theta_\bk}&-i\Delta_s e^{ij_z\theta_\bk}\\ \al_R(k_y-ik_x)&\xi_{\bk}&i\Delta_s e^{ij_z\theta_\bk}&\Delta_{t}e^{i(j_z+1)\theta_\bk}\\ \Delta_{t}e^{-i(j_z-1)\theta_\bk}&-i\Delta_s e^{-ij_z\theta_\bk}&-\xi_{\bk}&\al_R(k_y-ik_x)\\ i\Delta_s e^{-ij_z\theta_\bk}&\Delta_{t}e^{-i(j_z+1)\theta_\bk}&\al_R(k_y+ik_x)&-\xi_{\bk}\end{array}\right),
\end{eqnarray}
\end{widetext}
where $\xi_\bk=\frac{k^2}{2m}-E_F$, $\Delta_s=\Delta_{a+}-\Delta_{b+}$ and $\Delta_t=\Delta_{a+}+\Delta_{b+}$. The pairing term is
\begin{eqnarray}
&&\Delta_t e^{i(j_z-1)\theta_\bk}c^\dagger_{\bk\uparrow}c^\dagger_{-\bk\uparrow}+\Delta_t e^{i(j_z+1)\theta_\bk}c^\dagger_{\bk\downarrow}c^\dagger_{-\bk\downarrow}\nonumber\\&&-i\Delta_s e^{ij_z\theta_\bk}(c^\dagger_{\bk\uparrow}c^\dagger_{-\bk\downarrow} -c^\dagger_{\bk\downarrow}c^\dagger_{-\bk\uparrow})+h.c.,
\end{eqnarray}
which is consistent with Eq.(\ref{eq:pairing}).

\subsection{Topological invariant}
To see TRS breaking explicitly, we express $h(\bk)$ in terms of Dirac matrices,
\begin{eqnarray}
h(\bk)=\sum_{a=1}^5 d_a(\bk)\Gamma^a+\sum_{a<b=1}^5 d_{ab}(\bk)\Gamma^{ab}.
\end{eqnarray}
We choose the five Dirac matrices $\Gamma^a$, which anticommute with each other, to be
\begin{eqnarray}
\Gamma^a=(\sigma_z\otimes\mathbb{1},\sigma_x\otimes\mathbb{1},\sigma_y\otimes\vec{\bf{\sigma}}),
\end{eqnarray}
and
\begin{eqnarray}
\Gamma^{ab}=\frac{1}{2i}[\Gamma_a,\Gamma_b].
\end{eqnarray}
In this representation, the time reversal operator is $\mathcal{T}=(\mathbb{1}\otimes i\sigma_y)K$ where $K$ is the complex conjugate operator. The five Dirac matrices are even under time reversal, while the ten commutators are odd. If the coefficients satisfy
\begin{eqnarray}\label{eq:gi}
d_a(-\bk)=d_a(\bk), d_{ab}(-\bk)=-d_{ab}(\bk),
\end{eqnarray}
then the Hamiltonian is time-reversal invariant. However, the nine nonzero coefficients in $h(\bk)$ are
\begin{eqnarray}
d_1&=&\xi_\bk,\nonumber\\
d_2&=&\Delta_t\cos (j_z\theta_\bk)\cos\theta_\bk,\nonumber\\
d_4&=&-\Delta_s\sin (j_z\theta_\bk),\nonumber\\
d_5&=&\Delta_t\cos (j_z\theta_\bk)\sin\theta_\bk,\nonumber\\
d_{12}&=&-\Delta_t\sin (j_z\theta_\bk)\cos\theta_\bk,\nonumber\\
d_{14}&=&-\Delta_s\cos (j_z\theta_\bk),\nonumber\\
d_{15}&=&-\Delta_t\sin (j_z\theta_\bk)\sin\theta_\bk,\nonumber\\
d_{24}&=&-\al_R k_x,\nonumber\\
d_{45}&=&\al_R k_y.
\end{eqnarray}
Since $j_z$ is an even number, three terms out of the nine do not satisfy Eq.(\ref{eq:gi}), which are $d_2,d_5$ and $d_{14}$. Therefore, the Hamiltonian breaks TRS.

Similar to ($p_x+ip_y$)-wave superconductors which also break TRS, these states should be topologically nontrivial in the weak-pairing phase\cite{Volovik,ReadGreen}. Since in the previous calculations we assumed the chemical potential $\mu>0$, the states are always in the weak-pairing phase and topologically nontrivial. To see this explicitly, we calculate the Chern number of the system. The Chern number formula is \cite{BerryPhase}
\begin{eqnarray}\label{eq:Chern}
C=\frac{1}{2\pi}\sum_n\int d^2k\omega^n(\bk),
\end{eqnarray}
where the summation is over all occupied bands, and the Berry phase of the $n^{\rm{th}}$ band is
\begin{eqnarray}\label{eq:Berry}
\omega^n(\bk)=i\sum_{n'\ne n}\frac{\langle n|\frac{\partial h(\bk)}{\partial k_x}|n'\rangle\langle n'|\frac{\partial h(\bk)}{\partial k_y}|n\rangle-c.c.}{[\varepsilon_n(\bk)-\varepsilon_{n'}(\bk)]^2},
\end{eqnarray}
where $\varepsilon_n(\bk)$ and $|n\rangle$ are the $n^{\rm{th}}$ eigenenergy and eigenstate of $h(\bk)$, respectively. We have a four-band problem, two lower bands are occupied while two higher ones are empty, so the sum in Eq.(\ref{eq:Chern}) is over the two lower bands. There are four energy scales in the Hamiltonian, $\frac{k_R^2}{2m}$, $E_F$, $\Delta_s$ and $\Delta_t$. Similar as Eq.(\ref{eq:rescale}), we rescale the momentum $\bk$ by the Rashba momentum $k_R$, and then the Hamiltonian is rescaled by $\frac{k_R^2}{2m}$. Three dimensionless free parameters are left, which are $\frac{1}{\Theta}$, $\tilde{\Delta}_s=\frac{\Delta_s}{k_R^2/2m}$, and $\tilde{\Delta}_t=\frac{\Delta_t}{k_R^2/2m}$. The dimensionless Hamiltonian has the same Chern number as the original one, and is a function of the three parameters, $C(\frac{1}{\Theta},\tilde{\Delta}_s,\tilde{\Delta}_t)$. To evaluate Eq.(\ref{eq:Chern}) for a particular $j_z$, we have to resort to numerics. Although in principle the integral is over the whole $k$-space, the Berry curvature is negligible at large $k$ due to the large denominator in Eq.(\ref{eq:Berry}), so one can cut off the integral at a certain value of $k$. Furthermore, we need to convert the integral to Riemann sum over discrete points of a fine grid in $k$-space. As long as the gaps on both Fermi surfaces are not closed, the Chern number is quantized and does not change.

\begin{figure}[h]
\includegraphics[width=0.3\textwidth]{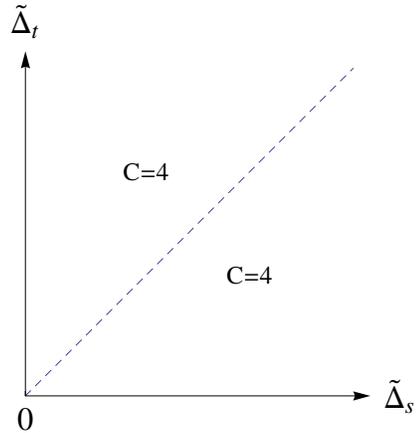}
\caption{The phase diagram of the topological superconducting states with $j_z=2$. The Chern number is 4 in the whole plane except the dashed line where $\tilde{\Delta}_s=\tilde{\Delta}_t$, where one of the gaps closes.} \label{fig:topological phase}
\end{figure}

We compute $C(\frac{1}{\Theta},\tilde{\Delta}_s,\tilde{\Delta}_t)$ for $j_z=2$ as an example. In practice, if the system is close to the transition point, i.e. the gap is very small, then the Berry curvature is highly peaked, and it is hard to make the Riemann sum converge. So we choose moderate gaps. For instance, for $\frac{1}{\Theta}\sim\mathcal{O}(1),\tilde{\Delta}_s\sim\mathcal{O}(0.1)$, and $\tilde{\Delta}_t\sim\mathcal{O}(0.01)$, the sum can be constrained in the region $|k_x|<3$ and $|k_y|<3$. If the interval between adjacent lines of the grid over which the sum is implemented is chosen to be 0.1, $C=4\pm0.2$; if the interval is 0.05, $C=4\pm0.01$. It converges to 4 as the interval becomes smaller and smaller.  We find that as long as both Fermi surfaces are gapped, i.e., both $\Delta_{a+}$ and $\Delta_{b+}$ are nonzero, or equivalently, $\tilde{\Delta}_s\ne\tilde{\Delta}_t$, the Chern number is $C=4$; if either gap or both gaps are closed, $C$ is not quantized. The phase diagram in terms of $\tilde{\Delta}_s$ and $\tilde{\Delta}_t$ is shown in Fig.\ref{fig:topological phase}. The phase diagram along the two axes is easily understood: if only $\Delta_s$ is finite, the system is like two copies of ($d+id$)-wave superconductors, thus $C=2+2=4$; if only $\Delta_t$ is finite, the system consists of a ($p+ip$)-wave superconductor and a ($f+if$)-wave superconductor, thus $C=1+3=4$. Physically, at large $\Theta$, $j_z=2$; as shown in Appendix \ref{mft}, $\Delta_{b+}$ is much smaller than $\Delta_{a+}$, and the sign of $\Delta_{b+}$ is determined by the sign of the Josephson coupling $V_{+-}^{(j_z=2)}$ which can be either positive or negative, so the state has $C=4$, and is near but can be either on the left or right of the black line in the phase diagram.

In general, $C=2j_z$ for the state with total angular momentum $j_z$.
\section{Summary}\label{summary}
In summary, we have studied the superconducting instability of a 2D repulsive Fermi gas with Rashba spin-orbit coupling. We implement a one-step renormalization group approach, and derive flow equations for the Cooper channel couplings in each angular momentum channel. The superconducting transition temperature $T_c$ is then identified with the highest energy scale of all the angular momentum channels at which the renormalized couplings diverge. We find that in general, $T_c$ increases with the dimensionless ratio $\Theta$, but there is an anomaly at $\Theta\sim 0.1$ where a dome appears. Starting from small $\Theta$, unconventional superconductivity occurs in a quite high angular momentum channel $j_z$, which decreases by a step 2 all the way to 2 as $\Theta$ increases, with an anomaly between $j_z=6$ and 4. In an extended range of $\Theta$, the superconducting gap predominately resides on the large Fermi surface, while momentum space Josephson coupling induces a smaller gap on the small Fermi surface. We develop a mean-field theory below $T_c$ and study the superconducting state. Self-consistent equations are derived and solved both at $T=0$ and just below $T_c$. In both cases, we find that the TRS breaking state, with full gaps on both Fermi surfaces, has the lowest condensation energy. The state with total angular momentum $j_z$ consists of singlets with orbital angular momentum $\ell=j_z$, spin-up triplets with $\ell=j_z-1$ and spin-down triplets with $\ell=j_z+1$. These chiral superconducting states are topologically nontrivial, and have nonzero Chern number $C=2j_z$.

\begin{acknowledgements}
We wish to thank Prof. L. P. Gor'kov for useful
discussions. L.W. would also like to thank Liang Sun for helpful discussions. This work is supported in part by NSF CAREER award
under Grant No. DMR-0955561.
\end{acknowledgements}
\appendix
\section{Mean field theory below $T_c$}\label{mft}
In this appendix, we apply mean field theory to the superconducting state below $T_c$ and solve the self-consistent equations. We find two solutions, of which one breaks TRS and the other does not. We then calculate the condensation energy at $T=0$ and just below $T_c$, and find that in both cases the TRS breaking state has a lower condensation energy.
\subsection{Self-consistent equations}
We start from the Hamiltonian
\begin{eqnarray}
H=H_{kin}+H_{int},
\end{eqnarray}
where
\begin{widetext}
\begin{eqnarray}
H_{kin}&=&\sum_{\bk,\lambda=\pm}(\eps_{\bk\lambda}-\mu)a^\dagger_{\bk\lambda} a_{\bk\lambda},\\
H_{int}&=&\frac{u^2m}{32L^2}\sum_{\bk\bk',\mu\lambda=\pm}e^{-i\theta_\bk}e^{i\theta_{\bk'}} \sum_{j_z=0,\pm2,\pm4...} V_{\mu\lambda}^{(j_z)}(e^{ij_z\theta_\bk}e^{-ij_z\theta_{\bk'}}+e^{-ij_z\theta_\bk}e^{ij_z\theta_{\bk'}}) a^\dagger_{\bk\mu}a^\dagger_{-\bk\mu}a_{-\bk'\lambda}a_{\bk'\lambda}.
\end{eqnarray}
\end{widetext}
Let $a_{\bk+}=a_\bk$ and $a_{\bk-}=b_{\bk}$. We replace the sum over all angular momentum channels with only the most dominant term, in which superconductivity occurs. Then
\begin{widetext}
\begin{eqnarray}
H_{int}&=&g_{++}\frac{1}{L^2}\sum_{s=\pm}\left(\sum_{\bk}e^{is(j_z-s)\theta_\bk}a^\dagger_{\bk}a^\dagger_{-\bk}\right) \left(\sum_{\bk'}e^{-is(j_z-s)\theta_{\bk'}}a_{-\bk'}a_{\bk'}\right)\nonumber\\
&+&g_{+-}\frac{1}{L^2}\sum_{s=\pm}\left(\sum_{\bk}e^{is(j_z-s)\theta_\bk}a^\dagger_{\bk}a^\dagger_{-\bk}\right) \left(\sum_{\bk'}e^{-is(j_z-s)\theta_{\bk'}}b_{-\bk'}b_{\bk'}\right)\nonumber\\
&+&g_{+-}\frac{1}{L^2}\sum_{s=\pm}\left(\sum_{\bk}e^{is(j_z-s)\theta_\bk}b^\dagger_{\bk}b^\dagger_{-\bk}\right) \left(\sum_{\bk'}e^{-is(j_z-s)\theta_{\bk'}}a_{-\bk'}a_{\bk'}\right),
\end{eqnarray}
\end{widetext}
where $g_{\mu\lambda}=\frac{u^2m}{32}V_{\mu\lambda}^{(j_z)}$. Now, let
\begin{eqnarray}
A_{j_z s}&=&\langle\frac{1}{L^2}\sum_{\bk}e^{-is(j_z-s)\theta_{\bk}}a_{-\bk}a_{\bk}\rangle,\\
B_{j_z s}&=&\langle\frac{1}{L^2}\sum_{\bk}e^{-is(j_z-s)\theta_{\bk}}b_{-\bk}b_{\bk}\rangle.
\end{eqnarray}
Neglecting the fluctuations, we have the mean field version of the interacting Hamiltonian
\begin{widetext}
\begin{eqnarray}
H_{int}&=&g_{++}\sum_{s=\pm}\left(A_{j_z s}\sum_{\bk}e^{is(j_z-s)\theta_\bk}a^\dagger_{\bk}a^\dagger_{-\bk} +A^*_{j_z s}\sum_{\bk}e^{-is(j_z-s)\theta_\bk}a_{-\bk}a_{\bk}-L^2 A^*_{j_z s}A_{j_z s}\right)\nonumber\\
&+&g_{+-}\sum_{s=\pm}\left(B_{j_z s}\sum_{\bk}e^{is(j_z-s)\theta_\bk}a^\dagger_{\bk}a^\dagger_{-\bk} +A^*_{j_z s}\sum_{\bk}e^{-is(j_z-s)\theta_\bk}b_{-\bk}b_{\bk}-L^2 A^*_{j_z s}B_{j_z s}\right)\nonumber\\
&+&g_{+-}\sum_{s=\pm}\left(A_{j_z s}\sum_{\bk}e^{is(j_z-s)\theta_\bk}b^\dagger_{\bk}b^\dagger_{-\bk} +B^*_{j_z s}\sum_{\bk}e^{-is(j_z-s)\theta_\bk}a_{-\bk}a_{\bk}-L^2 B^*_{j_z s}A_{j_z s}\right).
\end{eqnarray}
Summing only over half of the Brillouin zone, we have
\begin{eqnarray}\label{eq:MFH}
H&=&{\sum_\bk}'
 \left(\eps_{\bk+}-\mu+\eps_{\bk-}-\mu\right)\nonumber\\
&+&{\sum_{\bk}}'\left(a^\dagger_\bk,a_{-\bk}\right)\left(\begin{array}{cc}\eps_{\bk +}-\mu & 2\sum_{s=\pm}\Delta_{as}e^{i(sj_z-1)\theta_{\bk}}\\2\sum_{s=\pm}\Delta^*_{as}e^{-i(sj_z-1)\theta_{\bk}} & -\eps_{\bk +}+\mu \end{array}\right)\left(\begin{array}{cc}a_\bk\\a^\dagger_{-\bk} \end{array}\right)\nonumber\\
&+&{\sum_{\bk}}'\left(b^\dagger_\bk,b_{-\bk}\right)\left(\begin{array}{cc}\eps_{\bk -}-\mu & 2\sum_{s=\pm}\Delta_{bs}e^{i(sj_z-1)\theta_{\bk}}\\2\sum_{s=\pm}\Delta^*_{bs}e^{-i(sj_z-1)\theta_{\bk}} & -\eps_{\bk -}+\mu \end{array}\right)\left(\begin{array}{cc}b_\bk\\b^\dagger_{-\bk} \end{array}\right)\nonumber\\
&-&L^2\sum_{s=\pm}[g_{++}A^*_{j_z s}A_{j_z s}+g_{+-}(A^*_{j_z s}B_{j_z s}+B^*_{j_z s}A_{j_z s})],
\end{eqnarray}
\end{widetext}
where
\begin{eqnarray}
\Delta_{as}&=&g_{++}A_{j_z s}+g_{+-}B_{j_z s},\\
\Delta_{bs}&=&g_{+-}A_{j_z s}.
\end{eqnarray}
Now, let
\begin{eqnarray}\label{eq:gap}
\Delta_j(\bk)&=&\sum_{s=\pm}\Delta_{js}e^{i(sj_z-1)\theta_\bk}
\end{eqnarray}
where $j=a,b$.
The unitary transformation which diagonalizes the Hamiltonian is
\begin{eqnarray}
\left(\begin{array}{cc}u_{\bk j}&-v^*_{\bk j}\\v_{\bk j}&u^*_{\bk j}\end{array}\right)\left(\begin{array}{cc}\gamma_{\bk j}\\\gamma^\dagger_{-\bk j}\end{array}\right) =\left(\begin{array}{cc}j_{\bk}\\j^\dagger_{-\bk}\end{array}\right)
\end{eqnarray}
where the elements of the matrix satisfy
\begin{eqnarray}
\left(\begin{array}{cc}\eps_{\bk j}-\mu&2\Delta_j(\bk)\\2\Delta^*_j(\bk)&-\eps_{\bk j}+\mu\end{array}\right)\left(\begin{array}{cc}u_{\bk j}\\v_{\bk j}\end{array}\right)=E_{\bk j}\left(\begin{array}{cc}u_{\bk j}\\v_{\bk j}\end{array}\right).
\end{eqnarray}
Solving this equation, we get
\begin{eqnarray}\label{eq:disp}
E^2_{\bk j}=(\eps_{\bk j}-\mu)^2+4|\Delta_j(\bk)|^2,
\end{eqnarray}
where
\begin{eqnarray}
|\Delta_j(\bk)|^2=|\Delta_{j+}|^2+|\Delta_{j-}|^2 +2|\Delta_{j+}||\Delta_{j-}|\cos{(2j_z\theta_\bk+\al_{j})}.\nonumber\\
\end{eqnarray}
In the above equation, $\al_j$ is the difference between $\al_{j\pm}$, the phases of $\Delta_{j\pm}$, defined by
\begin{eqnarray}
\Delta_{j\pm}=|\Delta_{j\pm}|e^{i\al_{j\pm}}.
\end{eqnarray}
The eigenvectors of the matrix are
\begin{eqnarray}
\left(\begin{array}{cc}u_{\bk j}\\v_{\bk j}\end{array}\right)=\frac{1}{\sqrt{2}}\left(\begin{array}{cc}\frac{\Delta_j(\bk)}{|\Delta_j(\bk)|} \sqrt{1+\frac{\eps_{\bk j}-\mu}{E_{\bk j}}}\\\sqrt{1-\frac{\eps_{\bk j}-\mu}{E_{\bk j}}}\end{array}\right).
\end{eqnarray}
Then
\begin{widetext}
\begin{eqnarray}
A_{j_z s}&=&2\int'\frac{d^2\bk}{(2\pi)^2}e^{-i(sj_z-1)\theta_\bk}\langle(v^*_{\bk a}\gamma^\dagger_{\bk a}+u_{\bk a}\gamma_{-\bk a})(u_{\bk a}\gamma_{\bk a}-v^*_{\bk a}\gamma^\dagger_{-\bk a})\rangle,\nonumber\\
&=&-2\int'\frac{d^2\bk}{(2\pi)^2}e^{-i(sj_z-1)\theta_{\bk}}\frac{\Delta_a(\bk)}{E_{\bk a}}\left(1-2n_F(E_{\bk a})\right)\\
B_{j_z s}&=&-2\int'\frac{d^2\bk}{(2\pi)^2}e^{-i(sj_z-1)\theta_\bk}\frac{\Delta_b(\bk)}{E_{\bk b}}\left(1-2n_F(E_{\bk b})\right).
\end{eqnarray}
The self-consistent equations are
\begin{eqnarray}
\Delta_{as}&=&-2g_{++}\int'\frac{d^2\bk}{(2\pi)^2}e^{-i(sj_z-1)\theta_\bk}\frac{\Delta_{a+}e^{i(j_z-1)\theta_{\bk}}+ \Delta_{a-}e^{-i(j_z+1)\theta_{\bk}}}{E_{\bk a}}[1-2n_F(E_{\bk a})]\nonumber\\
&&-2g_{+-}\int'\frac{d^2\bk}{(2\pi)^2}e^{-i(sj_z-1)\theta_\bk}\frac{\Delta_{b+}e^{i(j_z-1)\theta_{\bk}}+ \Delta_{b-}e^{-i(j_z+1)\theta_{\bk}}}{E_{\bk b}}[1-2n_F(E_{\bk b})],\label{eq:sce1}\\
\Delta_{bs}&=&-2g_{+-}\int'\frac{d^2\bk}{(2\pi)^2}e^{-i(sj_z-1)\theta_\bk}\frac{\Delta_{a+}e^{i(j_z-1)\theta_{\bk}}+ \Delta_{a-}e^{-i(j_z+1)\theta_{\bk}}}{E_{\bk a}}[1-2n_F(E_{\bk a})]\label{eq:sce2},
\end{eqnarray}
\end{widetext}
where the integrals are over half of the Brillouin zone.
\subsection{Condensation energies at zero temperature}
The diagonalized Hamiltonian is
\begin{eqnarray}
H&=&{\sum_\bk}'(\eps_{\bk+}-\mu+\eps_{\bk-}-\mu)\nonumber\\
&+&{\sum_\bk}'\left(E_{\bk a}\gamma^\dagger_{\bk a}\gamma_{\bk a}-E_{\bk a}\gamma_{-\bk a}\gamma^\dagger_{-\bk a}\right)\nonumber\\
&+&{\sum_\bk}'\left(E_{\bk b}\gamma^\dagger_{\bk b}\gamma_{\bk b}-E_{\bk b}\gamma_{-\bk b}\gamma^\dagger_{-\bk b}\right)\nonumber\\
&-&L^2\sum_{s=\pm}[g_{++}A^*_{j_z s}A_{j_z s}+g_{+-}(A^*_{j_z s}B_{j_z s}+B^*_{j_z s}A_{j_z s})]\nonumber\\
\end{eqnarray}
The ground state energy of the condensate is then
\begin{eqnarray}
E_{gs}&=&{\sum_\bk}'(\eps_{\bk+}-\mu+\eps_{\bk-}-\mu)-{\sum_\bk}'(E_{\bk a}+E_{\bk b})\nonumber\\
&-&L^2\sum_{s=\pm}\left(\frac{\Delta^*_{as}\Delta_{bs}+\Delta^*_{bs}\Delta_{as}}{g_{+-}} -\frac{g_{++}}{g^2_{+-}}|\Delta_{bs}|^2\right)\nonumber\\
\end{eqnarray}
while the ground state energy of the normal state is
\begin{eqnarray}
E_0&=&\sum_{|\bk|<k_F}(\eps_{\bk +}-\mu+\eps_{\bk-}-\mu)\nonumber\\
&=&2{\sum_{|\bk|<k_F}}'(\eps_{\bk +}-\mu+\eps_{\bk-}-\mu).
\end{eqnarray}
Using self-consistent equations (\ref{eq:sce1}) and (\ref{eq:sce2}) with $n_F(E_{\bk a})=n_F(E_{\bk b})=0$ at $T=0$, we can write the the condensation energy as
\begin{eqnarray}
&&E_{gs}-E_0\nonumber\\
&=&-{\sum_{\bk}}'(E_{\bk a}-|\eps_{\bk+}-\mu|+E_{\bk b}-|\eps_{\bk-}-\mu|)\nonumber\\
&+&2{\sum_{\bk}}'\left[\frac{|\Delta_{a+}|^2+|\Delta_{a-}|^2+2|\Delta_{a+}||\Delta_{a-}|\cos(2j_z\theta_\bk+\al_a)} {E_{\bk a}}\right]\nonumber\\
&+&2{\sum_{\bk}}'\left[\frac{|\Delta_{b+}|^2+|\Delta_{b-}|^2+2|\Delta_{b+}||\Delta_{b-}|\cos(2j_z\theta_\bk+\al_b)} {E_{\bk b}}\right]\nonumber\\
\end{eqnarray}
Changing ${\sum_{\bk}}$ to $N_j\int_{-A}^A d\xi_j\int_0^{2\pi}\frac{d\theta_\bk}{2\pi}$ where $\xi_j=\eps_{\bk j}-\mu$ for $j=a,b$ and performing the integral, we have
\begin{eqnarray}
&&E_{gs}-E_0=\nonumber\\
&&-N_a(|\Delta_{a+}|^2+|\Delta_{a-}|^2)-N_b(|\Delta_{b+}|^2+|\Delta_{b-}|^2).
\end{eqnarray}
Now if $\frac{g_{+-}}{g_{++}}\ll 1$, the self-consistent equations can be solved iteratively by expanding
\begin{eqnarray}
\Delta_{as}&=&\Delta^{(0)}_{as}+\Delta^{(1)}_{as}+...\\
\Delta_{bs}&=&\Delta^{(1)}_{bs}+...
\end{eqnarray}
At the zeroth order, $g_{+-}$ is taken to be zero, and after integration the self-consistent equations become
\begin{widetext}
\begin{eqnarray}\label{eq:sce}
\frac{1}{g_{++}N_a}&=&-\ln\frac{4A^2}{|\Delta_{a+}|^2+|\Delta_{a-}|^2} -F_1\left(\frac{2|\Delta_{a+}||\Delta_{a-}|}{|\Delta_{a+}|^2+|\Delta_{a-}|^2}\right) -\frac{|\Delta_{a-}|}{|\Delta_{a+}|}F_2\left(\frac{2|\Delta_{a+}||\Delta_{a-}|}{|\Delta_{a+}|^2+ |\Delta_{a-}|^2}\right),\\
\frac{1}{g_{++}N_a}&=&-\ln\frac{4A^2}{|\Delta_{a+}|^2+|\Delta_{a-}|^2} -F_1\left(\frac{2|\Delta_{a+}||\Delta_{a-}|}{|\Delta_{a+}|^2+|\Delta_{a-}|^2}\right) -\frac{|\Delta_{a+}|}{|\Delta_{a-}|}F_2\left(\frac{2|\Delta_{a+}||\Delta_{a-}|}{|\Delta_{a+}|^2+ |\Delta_{a-}|^2}\right)
\end{eqnarray}
\end{widetext}
where
\begin{eqnarray}
F_1(\eta)&=&\int_0^{2\pi}\frac{d\theta_\bk}{2\pi}\ln\left(\frac{1}{1+\eta\cos(2j_z\theta_\bk)}\right)\nonumber\\ &=&-\ln\left(\frac{1}{2}+\frac{1}{2}\sqrt{1-\eta^2}\right),\\
F_2(\eta)&=&\int_0^{2\pi}\frac{d\theta_\bk}{2\pi}\cos(2j_z\theta_\bk)\ln\left(\frac{1}{1+\eta\cos(2j_z\theta_\bk)} \right) \nonumber\\ &=&\frac{\sqrt{1-\eta^2}-1}{\eta}
\end{eqnarray}
for any integer $j_z\neq 0$. There are two solutions to the self-consistent equations, either $|\Delta_{a+}|=|\Delta_{a-}|$ or $\Delta^{(0)}_{a+}\neq 0$ and $\Delta^{(0)}_{a-}=0$ (or vice versa). The first one breaks only rotational symmetry and the spectrum has nodes, while the second breaks TRS and the spectrum is gapped. The physical state is the one that has a lower condensation energy. In the first case,
\begin{eqnarray}
|\Delta_{a+}|^2=|\Delta_{a-}|^2=\frac{4A^2}{e}e^{\frac{1}{V}},
\end{eqnarray}
which yields condensation energy
\begin{eqnarray}
E_{gs}-E_0&=&-\frac{2}{e}N_a4A^2 e^{\frac{1}{g_{++}N_a}},
\end{eqnarray}
and for $|\Delta_{a+}|\neq 0$ and $|\Delta_{a-}|=0$ (or vice versa)
\begin{eqnarray}
|\Delta_{a+}|^2&=&4A^2 e^{\frac{1}{g_{++}N_a}},
\end{eqnarray}
then the condensation energy is
\begin{eqnarray}
E_{gs}-E_0&=&-N_a4A^2 e^{\frac{1}{g_{++}N_a}}.
\end{eqnarray}
Since $2/e<1$ the condensation energy is lower for the TRS breaking state with $|\Delta_{a+}|\neq 0$ and $|\Delta_{a-}|=0$.

To proceed with next order
\begin{eqnarray}
\Delta^{(1)}_{b+}=-2g_{+-}\int'\frac{d^2\bk}{(2\pi)^2}\frac{\Delta^{(0)}_{a+}+\Delta^{(0)}_{a-}e^{-2ij_z\theta_\bk}} {E_{\bk a}^{(0)}} =\frac{g_{+-}}{g_{++}}\Delta^{(0)}_{a+},\nonumber\\ \\
\Delta^{(1)}_{b-}=-2g_{+-}\int'\frac{d^2\bk}{(2\pi)^2}\frac{\Delta^{(0)}_{a+}e^{2ij_z\theta_\bk}+\Delta^{(0)}_{a-}} {E_{\bk a}^{(0)}} =\frac{g_{+-}}{g_{++}}\Delta^{(0)}_{a-}.\nonumber\\
\end{eqnarray}
This will correct the condensation energy to order $\frac{g^2_{+-}}{g^2_{++}}$.

Now, we need to find the correction to $\Delta_{as}$. Inspired by the fact that to first order in small $g_{+-}$, $\Delta_{bs}$ is proportional to $\Delta_{as}$, we seek a solution to the self-consistent equations, for arbitrary $g_{+-}$, in which we set $\Delta_{bs}=c\Delta_{as}$ where $c$ is some undetermined proportionality constant. Then, after performing the integral over the radial coordinate $\eps_{\bk s}-\mu$, we find
\begin{widetext}
\begin{eqnarray}
\Delta_{a+}&=&-g_{++}N_a\int_0^{2\pi}\frac{d\theta_\bk}{2\pi}(\Delta_{a+}+\Delta_{a-}e^{-2ij_z\theta_\bk}) \ln\frac{4A^2}{|\Delta_{a+}|^2+|\Delta_{a-}|^2+2|\Delta_{a+}||\Delta_{a-}|\cos(2j_z\theta_\bk+\al_a)}\nonumber\\
&-&g_{+-}cN_b\int_0^{2\pi}\frac{d\theta_\bk}{2\pi}(\Delta_{a+}+\Delta_{a-}e^{-2ij_z\theta_\bk}) \left[\ln\frac{1}{c^2}+\ln\frac{4A^2}{|\Delta_{a+}|^2+|\Delta_{a-}|^2+2|\Delta_{a+}||\Delta_{a-}|\cos(2j_z\theta_\bk +\al_a)} \right],\\
\Delta_{a-}&=&-g_{++}N_a\int_0^{2\pi}\frac{d\theta_\bk}{2\pi}(\Delta_{a+}e^{2ij_z\theta_\bk}+\Delta_{a-}) \ln\frac{4A^2}{|\Delta_{a+}|^2+|\Delta_{a-}|^2+2|\Delta_{a+}||\Delta_{a-}|\cos(2j_z\theta_\bk+\al_a)}\nonumber\\
&-&g_{+-}cN_b\int_0^{2\pi}\frac{d\theta_\bk}{2\pi}(\Delta_{a+}e^{2ij_z\theta_\bk}+\Delta_{a-}) \left[\ln\frac{1}{c^2}+\ln\frac{4A^2}{|\Delta_{a+}|^2+|\Delta_{a-}|^2+2|\Delta_{a+}||\Delta_{a-}|\cos(2j_z\theta_\bk +\al_a)} \right],\\
c\Delta_{a+}&=&-g_{+-}N_a\int_0^{2\pi}\frac{d\theta_\bk}{2\pi}(\Delta_{a+}+\Delta_{a-}e^{-2ij_z\theta_\bk}) \ln\frac{4A^2}{|\Delta_{a+}|^2+|\Delta_{a-}|^2+2|\Delta_{a+}||\Delta_{a-}|\cos(2j_z\theta_\bk+\al_a)},\\
c\Delta_{a-}&=&-g_{+-}N_a\int_0^{2\pi}\frac{d\theta_\bk}{2\pi}(\Delta_{a+}e^{2ij_z\theta_\bk}+\Delta_{a-}) \ln\frac{4A^2}{|\Delta_{a+}|^2+|\Delta_{a-}|^2+2|\Delta_{a+}||\Delta_{a-}|\cos(2j_z\theta_\bk+\al_a)}.
\end{eqnarray}
\end{widetext}
Comparing the 1st and the 3rd, as well as 2nd and the 4th equations we find the constraint on $c$ to be
\begin{eqnarray}
\frac{1+g_{+-}N_b c\ln\frac{1}{c^2}}{g_{++}N_a+g_{+-}N_b c}=\frac{c}{g_{+-}N_a}
\end{eqnarray}
or equivalently
\begin{eqnarray}
\frac{N_a}{N_b}G_{+-}\left(1+G_{+-}c\ln\frac{1}{c^2}\right)=cG_{++}+c^2 G_{+-}
\end{eqnarray}
where $G_{+-}=N_b g_{+-}$ and $G_{++}=N_a g_{++}$. This equation is easily solved numerically for $c$ as a function of $G_{+-}$. Nevertheless, analytically,
\begin{eqnarray}
c&=&\frac{N_a}{N_b}\frac{G_{+-}}{G_{++}}+... \ \mbox{for }|G_{+-}|\ll 1,\label{eq:c1}\\
c&=&\pm 1\ \ \ \ \ \ \ \ \mbox{for }G_{+-}\rightarrow \pm \infty.\label{eq:c2}
\end{eqnarray}
Note also that due to the symmetry $G_{+-}\rightarrow -G_{+-}$ and $c\rightarrow -c$, we only need the solution for positive $G_{+-}$.
Numerically, the result depends on the ratio of the density of states and the actual value of $G_{++}$, but schematically it rises linearly and then saturates to 1, meaning for large Josephson coupling, the gaps on the two Fermi surfaces are the same.
The condensation energy is then
\begin{eqnarray}\label{eq:cond}
E_{gs}-E_0&=&-N_a(|\Delta_{a+}|^2+|\Delta_{a-}|^2)-N_b(|\Delta_{b+}|^2+|\Delta_{b-}|^2)\nonumber\\
&=&-(N_a+c^2 N_b)(|\Delta_{a+}|^2+|\Delta_{a-}|^2)
\end{eqnarray}
The self-consistent equations are similar as Eq.(\ref{eq:sce}), but $\frac{1}{g_{++}N_a}$ is replaced by
\begin{eqnarray}
\frac{1}{V}=\frac{1+g_{+-}N_b c \ln\frac{1}{c^2}}{g_{++}N_a+g_{+-}N_b c}.
\end{eqnarray}
The two solutions remain the same except the above difference, so the TRS breaking state has a lower condensation energy.
\subsection{Ginzburg-Landau theory}\label{GL}
From self-consistency equations (\ref{eq:sce1}-\ref{eq:sce2}), we can derive the Ginzburg-Landau (GL) equations (without gradient terms). We need to evaluate the integrals such as
\begin{equation}
\int\frac{d^2\bk}{(2\pi)^2}\frac{\tanh{\frac{\be E_{\bk j}}{2}}}{E_{\bk j}}.
\end{equation}
Using
\begin{eqnarray}
\frac{\tanh\frac{\be x}{2}}{2x}=\frac{1}{\be}\sum_{\omega_n}\frac{1}{x+i\omega_n}\frac{1}{x-i\omega_n}
\end{eqnarray}
where $\omega_n=\frac{2n+1}{\be}\pi$ is the Matsubara frequency with an integer $n$ ranging from $-\infty$ to $\infty$, and changing $\int\frac{d^2\bk}{(2\pi)^2}$ to $N_j\int_{-A}^A d\xi_j\int_0^{2\pi}\frac{d\theta_\bk}{2\pi}$ where $\xi_j=\eps_{\bk j}-\mu$, we have another form of that integral
\begin{eqnarray}
&&N_j\int_{-A}^A d\xi_j\int_0^{2\pi}\frac{d\theta_\bk}{2\pi}\left[\frac{\tanh\frac{\be\xi_j}{2}}{\xi_j}\right.\nonumber\\
&&+\left.\frac{1}{\be} \sum_{\omega_n} \left(\frac{1}{E_{\bk j}+i\omega_n}\frac{1}{E_{\bk j}-i\omega_n}- \frac{1}{\xi_j+i\omega_n}\frac{1}{\xi_j-i\omega_n}\right)\right].\nonumber\\
\end{eqnarray}
The integral of the first term is
\begin{eqnarray}
\int_{-A}^A d\xi\frac{\tanh\frac{\be\xi}{2}}{\xi}\approx2\ln\frac{A}{T}.
\end{eqnarray}
After the radial integral, the Matsubara sum becomes
\begin{eqnarray}
\pi\sum_{\omega_n}\left(\frac{1}{\sqrt{\omega_n^2+4|\Delta_j|^2}}-\frac{1}{|\omega_n|}\right)\approx -2\pi|\Delta_j|^2\sum_{\omega_n}\frac{1}{|\omega_n|^3},
\end{eqnarray}
and the sum can be evaluated using
\begin{eqnarray}
\sum_{\omega_n}\frac{1}{|\omega_n|^3}=\frac{1}{(\pi T)^3}\frac{7}{4}\zeta(3).
\end{eqnarray}
The angular integral is left but easy to do. Let
\begin{eqnarray}
B=\frac{7\zeta(3)}{(\pi T)^2}.
\end{eqnarray}
Then we have the GL equations
\begin{widetext}
\begin{eqnarray}
-|\Delta_{a+}|\left(\frac{1}{g_{++}}+2N_a\ln\frac{A}{T}\right) -2|\Delta_{b+}|N_b\frac{|g_{+-}|}{g_{++}}\ln\frac{A}{T}+N_aB(|\Delta_{a+}|^3+2|\Delta_{a+}||\Delta_{a-}|^2)&& \nonumber\\
+N_b\frac{|g_{+-}|}{g_{++}} B\left(|\Delta_{b+}|^3+2|\Delta_{b+}||\Delta_{b-}|^2\right)&=&0,\label{eq:B1}\\
-|\Delta_{a-}|\left(\frac{1}{g_{++}}+2N_a\ln\frac{A}{T}\right) -2|\Delta_{b-}|N_b\frac{|g_{+-}|}{g_{++}}\ln\frac{A}{T}+N_aB(|\Delta_{a-}|^3+2|\Delta_{a-}||\Delta_{a+}|^2)&& \nonumber\\ +N_b\frac{|g_{+-}|}{g_{++}} B\left(|\Delta_{b-}|^3+2|\Delta_{b-}||\Delta_{b+}|^2\right)&=&0,\label{eq:B2}
\end{eqnarray}
\begin{eqnarray}
-\frac{1}{|g_{+-}|}|\Delta_{b+}|-2|\Delta_{a+}|N_a\ln\frac{A}{T}+N_aB\left(|\Delta_{a+}|^3+2|\Delta_{a+}| |\Delta_{a-}|^2\right) &=&0 \label{eq:B3},\\
-\frac{1}{|g_{+-}|}|\Delta_{b-}|-2|\Delta_{a-}|N_a\ln\frac{A}{T}+N_aB\left(|\Delta_{a-}|^3+2|\Delta_{a-}| |\Delta_{a+}|^2\right) &=&0, \label{eq:B4}
\end{eqnarray}
where we have used the fact that the relative sign between $\Delta_{bs}$ and $\Delta_{as}$, $e^{i(\al_{as}-\al_{bs})}$, is the same the sign of $g_{+-}$, as in Eq.(\ref{eq:c1}-\ref{eq:c2}).
Subtracting (\ref{eq:B3}) from (\ref{eq:B1}) and (\ref{eq:B4}) from (\ref{eq:B2}), and multiplying $g_{++}/|g_{+-}|$, we get
\begin{eqnarray}
-\frac{1}{|g_{+-}|}|\Delta_{a+}|+\frac{g_{++}}{g_{+-}^2}|\Delta_{b+}|-2|\Delta_{b+}|N_b\ln\frac{A}{T}+N_b B\left(|\Delta_{b+}|^3+2|\Delta_{b+}||\Delta_{b-}|^2\right)=0,\label{eq:B5}\\
-\frac{1}{|g_{+-}|}|\Delta_{a-}|+\frac{g_{++}}{g_{+-}^2}|\Delta_{b-}|-2|\Delta_{b-}|N_b\ln\frac{A}{T}+N_b B\left(|\Delta_{b-}|^3+2|\Delta_{b-}||\Delta_{b+}|^2\right)=0.\label{eq:B6}
\end{eqnarray}
The four equations (\ref{eq:B3}-\ref{eq:B6}) can be regarded as the variational derivatives of the GL function with respect to $|\Delta_{a+}|,|\Delta_{a-}|,|\Delta_{b+}|$ and $|\Delta_{b-}|$, respectively. Therefore, we can deduce the GL function
\begin{eqnarray}
F&=&-\frac{1}{|g_{+-}|}(|\Delta_{a+}||\Delta_{b+}|+|\Delta_{a-}||\Delta_{b-}|)+\frac{1}{2}\frac{g_{++}}{g_{+-}^2} (|\Delta_{b+}|^2+|\Delta_{b-}|^2) -\ln\frac{A}{T}\left[N_a(|\Delta_{a+}|^2+|\Delta_{a-}|^2)+N_b(|\Delta_{b+}|^2+|\Delta_{b-}|^2)\right]\nonumber\\ &+&B\left[N_a(\frac{1}{4}|\Delta_{a+}|^4+\frac{1}{4}|\Delta_{a-}|^4+|\Delta_{a+}|^2|\Delta_{a-}|^2) +N_b(\frac{1}{4}|\Delta_{b+}|^4+\frac{1}{4}|\Delta_{b-}|^4+|\Delta_{b+}|^2|\Delta_{b-}|^2)\right].
\end{eqnarray}
Let $R_{j}=\sqrt{|\Delta_{j+}|^2+|\Delta_{j-}|^2}$ and $\theta_{j}=\tan^{-1}\frac{|\Delta_{j-}|}{|\Delta_{j+}|}$, or equivalently $|\Delta_{j+}|=R_{j}\cos\theta_{j},|\Delta_{j-}|=R_{j}\sin\theta_{j}$, where $\theta_j\in[0,\frac{\pi}{2}]$, then
\begin{eqnarray}
F=-\frac{1}{|g_{+-}|}R_aR_b\cos(\theta_a-\theta_b)+\frac{1}{2}\frac{g_{++}}{g_{+-}^2}R_b^2 -\ln\frac{A}{T}\left(N_aR_a^2+N_bR_b^2\right) +\frac{1}{4}B\left(N_aR_a^4(1+\frac{1}{2}\sin^22\theta_a)+N_bR_b^4(1+\frac{1}{2}\sin^22\theta_b)\right).\nonumber\\
\end{eqnarray}
For any positive $R_a$ and $R_b$, to minimize $F$, we need $\theta_a=\theta_b=0\mbox{ or }\frac{\pi}{2}$, which corresponds to the TRS breaking state found in last section. The other solution is $\Delta_{j+}=\Delta_{j-}$, or equivalently $\theta_a=\theta_b=\frac{\pi}{4}$, corresponds to the maximum of $F$. The coefficient of the quartic term is 1.5 times larger in the latter case than in the former case. In the TRS breaking state,
\begin{eqnarray}
F=-\frac{1}{|g_{+-}|}R_aR_b+\frac{1}{2}\frac{g_{++}}{g_{+-}^2}R_b^2-\ln\frac{A}{T}\left(N_aR_a^2+N_bR_b^2\right) +\frac{1}{4}B\left(N_aR_a^4+N_bR_b^4\right).
\end{eqnarray}
\end{widetext}
To find $T_c$, consider only the quadratic term in $F$,
\begin{eqnarray}
F'=-\frac{1}{2}\left[R_a,R_b\right]\left[\begin{array}{cc}2N_a\ln\frac{A}{T}&\frac{1}{|g_{+-}|} \\\frac{1}{|g_{+-}|}&2N_b\ln\frac{A}{T} -\frac{g_{++}}{|g_{+-}|^2}\end{array}\right] \left[\begin{array}{c}R_a\\R_b\end{array}\right]\nonumber\\
\end{eqnarray}
and $T_c$ is obtained by setting $F'=0$. In terms of $T_c$, the GL function can be written as
\begin{eqnarray}
F=\mathcal{A}\left(N_aR_a^2+N_bR_b^2\right)+\frac{1}{4}B\left(N_aR_a^4+N_bR_b^4\right)
\end{eqnarray}
where $\mathcal{A}=\ln\frac{T}{T_c}\approx (T-T_c)/T$. The minimum of $F$ is at
\begin{eqnarray}
R_a^2=R_b^2=-\frac{2\mathcal{A}}{B},
\end{eqnarray}
and the condensation energy is
\begin{eqnarray}
F_{min}=-\frac{\mathcal{A}^2}{B}(N_a+N_b).
\end{eqnarray}
Obviously, this is 1.5 times lower than in the case without TRS breaking where $B$ is replaced by $1.5B$.
\bibliography{SObib}

\begin{thebibliography}{15}
\expandafter\ifx\csname natexlab\endcsname\relax\def\natexlab#1{#1}\fi
\expandafter\ifx\csname bibnamefont\endcsname\relax
  \def\bibnamefont#1{#1}\fi
\expandafter\ifx\csname bibfnamefont\endcsname\relax
  \def\bibfnamefont#1{#1}\fi
\expandafter\ifx\csname citenamefont\endcsname\relax
  \def\citenamefont#1{#1}\fi
\expandafter\ifx\csname url\endcsname\relax
  \def\url#1{\texttt{#1}}\fi
\expandafter\ifx\csname urlprefix\endcsname\relax\def\urlprefix{URL }\fi
\providecommand{\bibinfo}[2]{#2}
\providecommand{\eprint}[2][]{\url{#2}}

\bibitem[{\citenamefont{Kohn and Luttinger}(1965)}]{KohnLuttinger}
\bibinfo{author}{\bibfnamefont{W.}~\bibnamefont{Kohn}} \bibnamefont{and}
  \bibinfo{author}{\bibfnamefont{J.~M.} \bibnamefont{Luttinger}},
  \bibinfo{journal}{Phys. Rev. Lett.} \textbf{\bibinfo{volume}{15}},
  \bibinfo{pages}{524} (\bibinfo{year}{1965}).

\bibitem[{\citenamefont{Chubukov}(1993)}]{ChubukovPRB}
\bibinfo{author}{\bibfnamefont{A.~V.} \bibnamefont{Chubukov}},
  \bibinfo{journal}{Phys. Rev. B} \textbf{\bibinfo{volume}{48}},
  \bibinfo{pages}{1097} (\bibinfo{year}{1993}).

\bibitem[{\citenamefont{Edelshtein}(1989)}]{Edelstein}
\bibinfo{author}{\bibfnamefont{V.~M.} \bibnamefont{Edelshtein}},
  \bibinfo{journal}{Sov. Phys. JETP} \textbf{\bibinfo{volume}{68}},
  \bibinfo{pages}{1244} (\bibinfo{year}{1989}).

\bibitem[{\citenamefont{Gor'kov and Rashba}(2001)}]{GorkovRashba}
\bibinfo{author}{\bibfnamefont{L.~P.} \bibnamefont{Gor'kov}} \bibnamefont{and}
  \bibinfo{author}{\bibfnamefont{E.~I.} \bibnamefont{Rashba}},
  \bibinfo{journal}{Phys. Rev. Lett.} \textbf{\bibinfo{volume}{87}},
  \bibinfo{pages}{037004} (\bibinfo{year}{2001}).

\bibitem[{\citenamefont{{Mineev} and {Sigrist}}(2009)}]{MineevSigrist}
\bibinfo{author}{\bibfnamefont{V.~P.} \bibnamefont{{Mineev}}} \bibnamefont{and}
  \bibinfo{author}{\bibfnamefont{M.}~\bibnamefont{{Sigrist}}},
  \bibinfo{journal}{ArXiv e-prints}  (\bibinfo{year}{2009}),
  \eprint{0904.2962}.

\bibitem[{\citenamefont{Vafek and Wang}(2011)}]{VafekWang}
\bibinfo{author}{\bibfnamefont{O.}~\bibnamefont{Vafek}} \bibnamefont{and}
  \bibinfo{author}{\bibfnamefont{L.}~\bibnamefont{Wang}},
  \bibinfo{journal}{Phys. Rev. B} \textbf{\bibinfo{volume}{84}},
  \bibinfo{pages}{172501} (\bibinfo{year}{2011}),
  \urlprefix\url{http://link.aps.org/doi/10.1103/PhysRevB.84.172501}.

\bibitem[{\citenamefont{Shankar}(1994)}]{Shankar}
\bibinfo{author}{\bibfnamefont{R.}~\bibnamefont{Shankar}},
  \bibinfo{journal}{Reviews of Modern Physics} \textbf{\bibinfo{volume}{66}},
  \bibinfo{pages}{129} (\bibinfo{year}{1994}).

\bibitem[{\citenamefont{Gor'kov and
  Melik-Barkhudarov}(1961)}]{GorkovMelikBarkhudarov}
\bibinfo{author}{\bibfnamefont{L.~P.} \bibnamefont{Gor'kov}} \bibnamefont{and}
  \bibinfo{author}{\bibfnamefont{T.~K.} \bibnamefont{Melik-Barkhudarov}},
  \bibinfo{journal}{Sov. Phys. JETP} \textbf{\bibinfo{volume}{13}},
  \bibinfo{pages}{1018} (\bibinfo{year}{1961}).

\bibitem[{\citenamefont{Baranov et~al.}(1992)\citenamefont{Baranov, Chubukov,
  and Kagan}}]{ChubukovReview}
\bibinfo{author}{\bibfnamefont{M.~A.} \bibnamefont{Baranov}},
  \bibinfo{author}{\bibfnamefont{A.~V.} \bibnamefont{Chubukov}},
  \bibnamefont{and} \bibinfo{author}{\bibfnamefont{M.~Y.} \bibnamefont{Kagan}},
  \bibinfo{journal}{International Journal of Modern Physics B}
  \textbf{\bibinfo{volume}{6}}, \bibinfo{pages}{2471} (\bibinfo{year}{1992}).

\bibitem[{\citenamefont{Raghu et~al.}(2010)\citenamefont{Raghu, Kivelson, and
  Scalapino}}]{RaghuKivelsonScalapino}
\bibinfo{author}{\bibfnamefont{S.}~\bibnamefont{Raghu}},
  \bibinfo{author}{\bibfnamefont{S.~A.} \bibnamefont{Kivelson}},
  \bibnamefont{and} \bibinfo{author}{\bibfnamefont{D.~J.}
  \bibnamefont{Scalapino}}, \bibinfo{journal}{Phys. Rev. B}
  \textbf{\bibinfo{volume}{81}}, \bibinfo{pages}{224505}
  (\bibinfo{year}{2010}).

\bibitem[{\citenamefont{Raghu and Kivelson}(2011)}]{RaghuKivelson2011}
\bibinfo{author}{\bibfnamefont{S.}~\bibnamefont{Raghu}} \bibnamefont{and}
  \bibinfo{author}{\bibfnamefont{S.~A.} \bibnamefont{Kivelson}},
  \bibinfo{journal}{Phys. Rev. B} \textbf{\bibinfo{volume}{83}},
  \bibinfo{pages}{094518} (\bibinfo{year}{2011}).

\bibitem[{\citenamefont{Anderson and Morel}(1961)}]{AndersonMorel1961}
\bibinfo{author}{\bibfnamefont{P.~W.} \bibnamefont{Anderson}} \bibnamefont{and}
  \bibinfo{author}{\bibfnamefont{P.}~\bibnamefont{Morel}},
  \bibinfo{journal}{Phys. Rev.} \textbf{\bibinfo{volume}{123}},
  \bibinfo{pages}{1911} (\bibinfo{year}{1961}).

\bibitem[{\citenamefont{Volovik}(2003)}]{Volovik}
\bibinfo{author}{\bibfnamefont{G.}~\bibnamefont{Volovik}},
  \emph{\bibinfo{title}{The Universe in a Helium Droplet}}
  (\bibinfo{publisher}{Clarendon Press, Oxford}, \bibinfo{year}{2003}).

\bibitem[{\citenamefont{Read and Green}(2000)}]{ReadGreen}
\bibinfo{author}{\bibfnamefont{N.}~\bibnamefont{Read}} \bibnamefont{and}
  \bibinfo{author}{\bibfnamefont{D.}~\bibnamefont{Green}},
  \bibinfo{journal}{Phys. Rev. B} \textbf{\bibinfo{volume}{61}},
  \bibinfo{pages}{10267} (\bibinfo{year}{2000}).

\bibitem[{\citenamefont{Xiao et~al.}(2010)\citenamefont{Xiao, Chang, and
  Niu}}]{BerryPhase}
\bibinfo{author}{\bibfnamefont{D.}~\bibnamefont{Xiao}},
  \bibinfo{author}{\bibfnamefont{M.-C.} \bibnamefont{Chang}}, \bibnamefont{and}
  \bibinfo{author}{\bibfnamefont{Q.}~\bibnamefont{Niu}}, \bibinfo{journal}{Rev.
  Mod. Phys.} \textbf{\bibinfo{volume}{82}}, \bibinfo{pages}{1959}
  (\bibinfo{year}{2010}).

\end{thebibliography}
\end{document}